\documentclass[10pt,conference]{IEEEtran}
\usepackage{cite}
\usepackage{amsmath,amssymb,amsfonts}
\usepackage{algorithmic}
\usepackage{graphicx}
\ifdefined\pdfdraftmode
\fi
\usepackage{textcomp}
\usepackage{xcolor}
\usepackage[hyphens]{url}
\usepackage{fancyhdr}
\usepackage[hidelinks,hyperfootnotes=false]{hyperref}
\usepackage{array}
\usepackage{tabularx}
\usepackage{multirow}
\usepackage{colortbl}
\usepackage{etoolbox}
\title{Beyond Capacity: Scalable MoE LLM Inference via High-Bandwidth Flash with Direct GPU and HBM Paths}

\author{
  \IEEEauthorblockN{
    Seeyeon Kim$^{*}$, Juhyeong Jin$^{*}$ and Joo-Young Kim
  }
  \IEEEauthorblockA{
    KAIST \\
    Daejeon, Republic of Korea \\
    \{seeyakim, pearlbro, jooyoung1203\}@kaist.ac.kr
  }
}
\newcommand{\systemname}{DASH}

\begin{document}
\maketitle
\thispagestyle{plain}
\pagestyle{plain}
\begingroup
\renewcommand{\thefootnote}{\fnsymbol{footnote}}
\footnotetext[1]{Both authors contributed equally to this research.}
\endgroup

\makeatletter
\let\DASHorigmakecaption\@makecaption
\long\def\@makecaption#1#2{%
  \begingroup
    \let\footnotesize\small
    \DASHorigmakecaption{#1}{#2}%
  \endgroup
}
\makeatother


\begin{abstract}
    Modern mixture-of-experts (MoE) language models increasingly strain the capacity and cost efficiency of high-bandwidth memory (HBM), as rapidly growing expert weights must be provisioned close to GPUs.
    High-bandwidth flash (HBF) offers substantially greater capacity, but conventional designs typically deliver HBF-resident expert weights to the GPU through HBM, leaving an additional direct GPU–HBF connection underutilized.
    We explore an HBF organization that simultaneously exploits two independent expert-delivery routes: a direct path that transfers expert weights from HBF to the GPU and a relay path that transfers them from HBF through the HBM base die to the GPU.
    Whole experts are assigned to one of the two routes, and transfers over both routes proceed concurrently, increasing aggregate expert-delivery bandwidth without replicating expert weights or introducing a shared relay bottleneck.
    Early expert determination identifies upcoming experts ahead of their conventional execution point, allowing HBF read latency to overlap with preceding computation, while separate management of immutable expert weights and mutable KV-cache data reduces interference between the two traffic classes.
    We evaluate the architecture using an event-driven continuous-batching LLM serving simulator with empirically measured GPU compute latencies.
    Across representative MoE workloads, concurrently utilizing the direct GPU–HBF and HBF–HBM–GPU routes consistently improves expert-delivery efficiency over designs restricted to either route alone.
    For a representative workload, the proposed architecture can achieve 1.94$\times$ higher throughput and 1.90$\times$ end-to-end speedup over a design that delivers all HBF-resident expert weights to the GPU through the HBM base die.
\end{abstract}

\section{Introduction}
\label{sec:introduction}

\begin{figure}[t]
  \centering
  \includegraphics[width=\columnwidth]{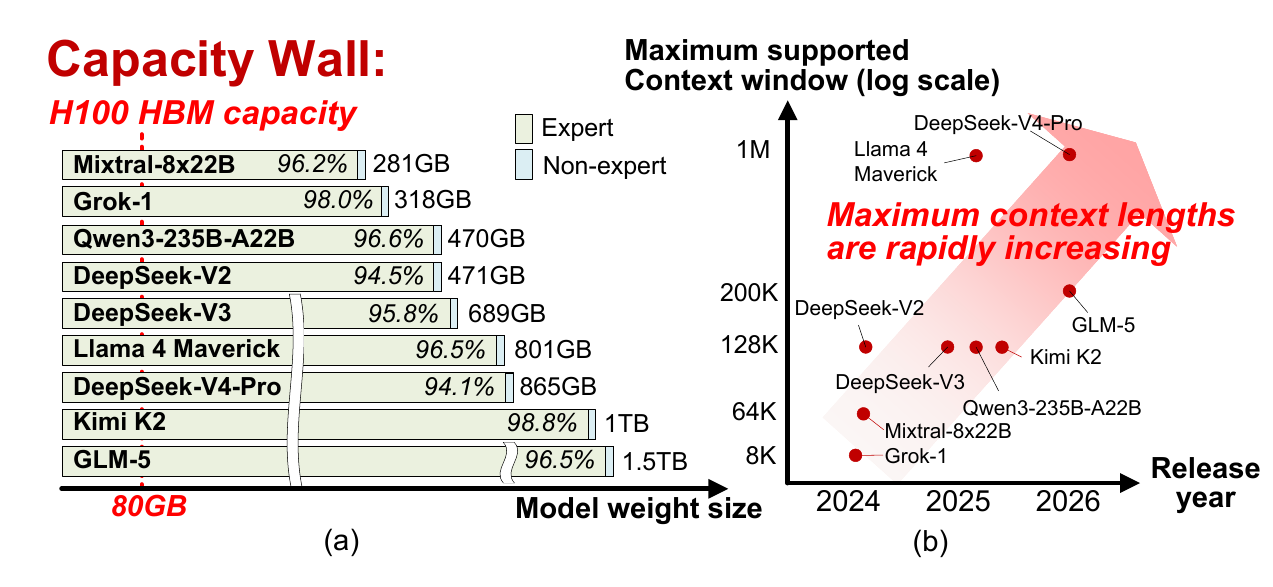}
   \caption{(a) Model weight sizes and expert-weight fractions and (b) supported context-window lengths, derived from the published model configurations~\cite{deepseekv2,deepseekv3,deepseekv4,llama4_maverick,mixtral8x22b,qwen3_235b,kimi_k2,glm5,grok1}.}
  \label{fig:fig_1}
\end{figure}

Modern LLM inference puts rapidly growing pressure on memory capacity and bandwidth as model sizes and context lengths increase.
MoE models increase total parameter capacity by incorporating many experts, while limiting per-token computation to a small subset of activated experts~\cite{MoE, shazeer2017moe,fedus2022switch}.
Meanwhile, long-context requests generate larger key--value (KV) caches that must be retained throughout decoding~\cite{deepseekv4,llama4_maverick}.
Continuous batching interleaves the prefill and decode phases of multiple requests, causing their KV caches to coexist in memory~\cite{yu2022orca,sarathi_serve}.
These requests also invoke MoE layers with token-dependent expert selections, requiring the full expert set to remain stored and the selected weights to be delivered at each step.
Together, these trends increase two major components of the memory footprint: model weights grow with the total parameter count, whereas KV cache size grows with context length and batch size.
Figure~\ref{fig:fig_1}(a) illustrates this capacity pressure, showing that the surveyed MoE checkpoints have weight footprints of 281\,GB--1.5\,TB, with expert weights accounting for 94.1--98.8\% of the total weight.
Each model's weights alone exceed the NVIDIA H100's 80\,GB HBM capacity~\cite{nvidia_h100}, even before accounting for KV caches or activations.
Figure~\ref{fig:fig_1}(b) shows supported context windows reaching hundreds of thousands to millions of tokens, making KV cache and expert weight an increasingly significant memory component under long contexts and high concurrency.

High Bandwidth Flash (HBF)~\cite{sandisk_hbf} offers a promising approach to alleviating this \textbf{capacity wall}.
HBF is a die-stacked NAND memory that combines high capacity with aggregate read bandwidth from concurrent accesses across multiple dies and planes.
Prior works~\cite{h3, hbm_hbf_pooling} have explored integrating HBF into the GPU memory hierarchy to leverage its capacity and bandwidth potential.
However, such approaches retain an HBM-centric architecture, with HBF primarily serving as a backing or capacity-extension tier.
Its cascaded GPU--HBM--HBF path requires HBF traffic to traverse HBM-side routing resources before reaching the GPU.
Consequently, HBF traffic shares the HBM-side delivery path, preventing HBF bandwidth from being exposed to the GPU as an independent resource.

Direct GPU access, however, leaves two NAND-induced latency challenges.
First, each read must sense a page from the NAND array into a page buffer before data transfer begins, incurring a startup latency of $t_R$~\cite{kioxia_tc58_2019,sandisk_hbf_patent_2025}.
For dense layers, weight accesses follow a deterministic execution order and can therefore be prefetched~\cite{h3}.
In MoE models, however, expert selection is input-dependent, and the selected experts are not known until routing completes, making conventional prefetching ineffective.
Prior work~\cite{hbm_hbf_pooling} avoids this problem by retaining expert weights in HBM, but the expert weights of recent large-scale MoE models~\cite{qwen3_235b,mixtral8x22b,grok1,llama4_maverick,deepseekv2,deepseekv3} can exceed the available HBM capacity.

Second, HBF program operations are much slower than reads and can delay latency-critical reads~\cite{kioxia_tc58_2019,agrawal2008ssd}.
Prefill generates large KV writes that can be programmed directly, whereas decode produces small updates that must be buffered and coalesced into pages.

\begin{figure}[t]
    \centering
    \includegraphics[width=\linewidth]{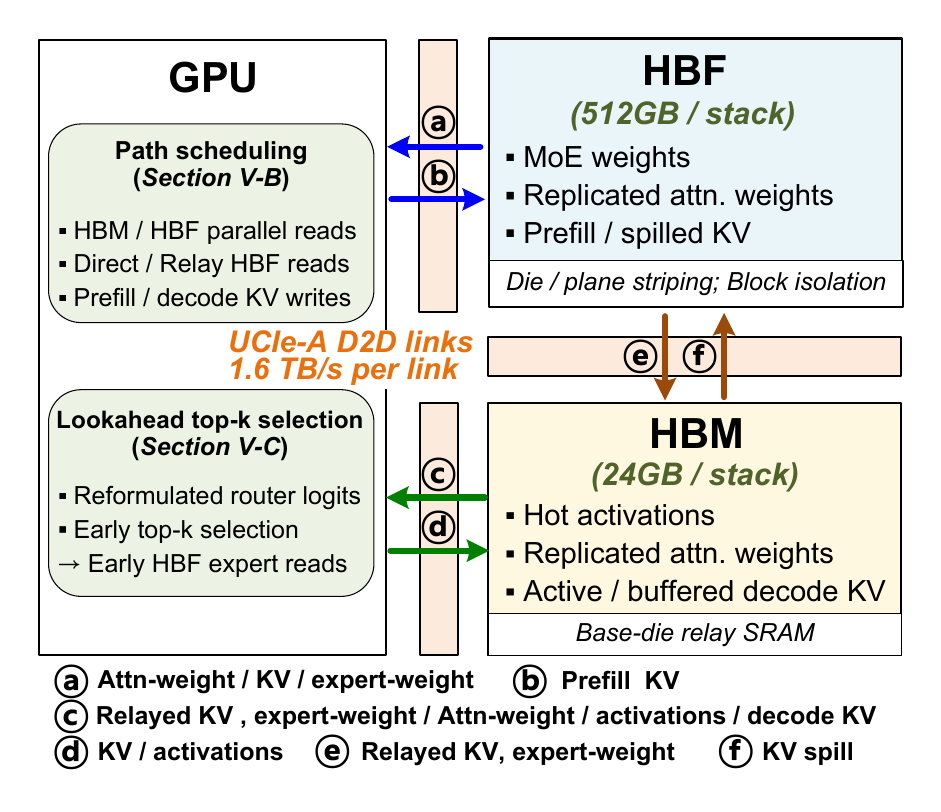}
    \caption{Overview of \systemname{}.}
    \label{fig:fig_2}
\end{figure}

To address these issues, we propose \textbf{\systemname{}} (\textbf{D}irect \textbf{A}ttachment of HBF to the GPU as the main memory tier, with a \textbf{S}eparate path to \textbf{H}BM), as illustrated in Figure~\ref{fig:fig_2}.
\systemname{} attaches both HBM and HBF to the GPU through independent UCIe links~\cite{sharma2022universal}, allowing each memory to serve GPU requests without relying on the other.
\systemname{} also connects the two memory base dies through a dedicated HBM--HBF path.
This path supports HBM-to-HBF writeback and relays HBF-resident reads through the HBM base die without accessing memory cells.
Thus, the direct GPU–HBF path and the HBF–HBM-base-die–GPU relay path can operate concurrently, increasing delivery bandwidth for HBF-resident data such as MoE expert weights.

To exploit these paths for large MoE LLM inference, \systemname{} places data across HBM and HBF according to its footprint, access pattern, and write behavior.
Frequently updated data and intermediate activations are retained in HBM, whereas large read-mostly data is placed in HBF.
Smaller non-expert weights reused by every token, such as attention QKV and output projection weights, are replicated across HBM and HBF.
Within HBF, model weights and KV cache are distributed across multiple dies and planes to expose the available parallelism.
We also separate model weights from KV cache at erase block granularity, preventing KV garbage collection from relocating model weight pages.
\systemname{} determines the top-$k$ experts early and initiates their HBF reads before conventional routing completes, hiding $t_R$.
Large prefill KV is written directly to HBF, whereas small decode updates are buffered and coalesced in HBM to hide program latency, $t_{\mathrm{PROG}}$.
 
Together, these mechanisms match HBF access to the distinct bandwidth, timing and write-granularity demands of sparse expert delivery under dynamic serving.
Therefore, \systemname{} elevates HBF beyond a capacity-extension tier, making it a first-class component of the GPU memory system for large-scale MoE inference.
With a batch size of 4, a 1K-token prefill, and 128 decode tokens, \systemname{} delivers five-model geometric-mean throughput speedup of $1.90\times$ and reduces end-to-end latency by 44.8\% over a cascaded-only baseline (RelayOnly), which delivers HBF data to the GPU through the HBM base-die relay path.
Under continuous batching at 75\% of this baseline's measured saturation rate, \systemname{} reduces Qwen3 P90 end-to-end latency by 53.5\%, compared with 53.3\% against a direct GPU--HBF baseline (DirectOnly).
In summary, this paper makes the following contributions:

\begin{itemize}
  \item We elevate HBF from a passive capacity extension to a main memory tier by combining direct GPU attachment with a dedicated HBM--HBF path, enabling concurrent dual-path delivery of HBF-resident data.

  \item We hide HBF access latency by initiating selected expert reads as soon as lookahead selection completes and scheduling KV writes according to their production and reuse timing.

  \item We integrate \systemname{} into a continuous-batching serving simulator that co-schedules decode tokens and chunked prefills at iteration boundaries, improving throughput and tail latency under dynamic arrivals.

  \item We establish \systemname{}’s system-level scalability, combining robust capacity-per-cost gains with a modular HBM–HBF topology.
  
\end{itemize}

\section{Background}
\label{sec:background}

\subsection{Modern MoE LLM Inference}
\label{sec:bg-moe-inference}

\begin{figure}[t]
    \centering
    \includegraphics[width=\linewidth]{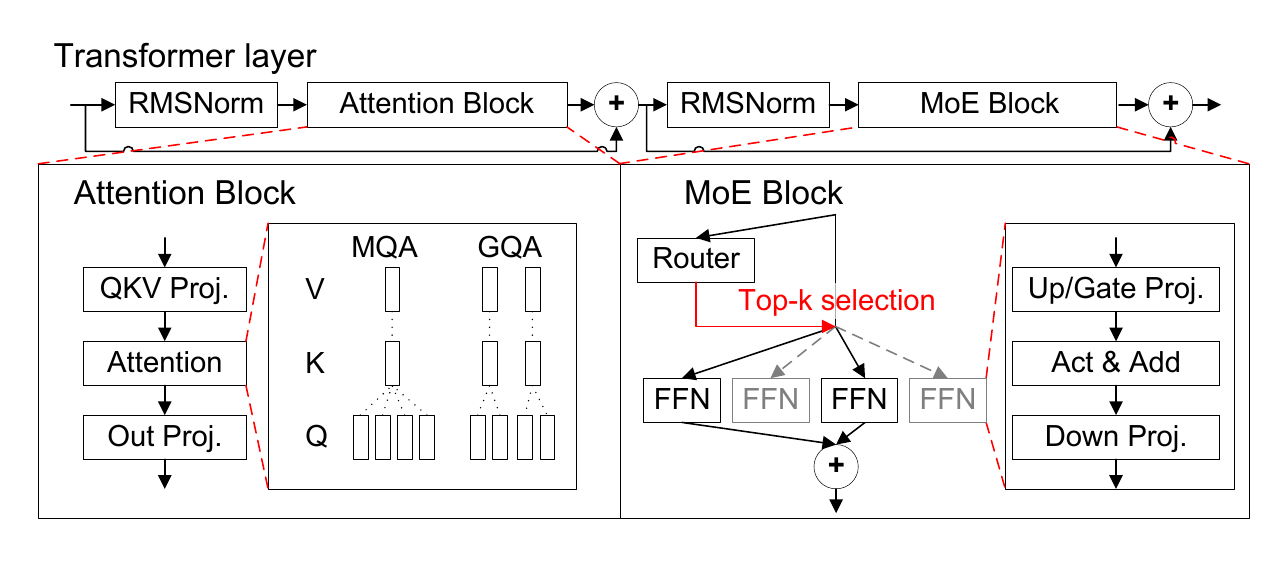}
    \caption{Structure of MoE Transformer layer}
    \label{fig:fig_3}
\end{figure}

\subsubsection{Transformer Layer}
Figure~\ref{fig:fig_3} shows an MoE Transformer layer~\cite{vaswani2017attention,shazeer2017moe,fedus2022switch}, which consists of an attention block followed by an MoE block, with RMS normalization~\cite{rmsnorm2019} and residual connections~\cite{he2016deep}.
When processing individual requests, MoE inference combines two distinct memory behaviors.
The attention block~\cite{vaswani2017attention} performs QKV and output projections while reading and updating the KV cache, whose capacity grows with context length despite the reduced number of KV heads in GQA~\cite{gqa} and MQA~\cite{mqa}.
The MoE block~\cite{shazeer2017moe,fedus2022switch} activates only the top-$k$ experts for each token, reducing computation but not the capacity required to keep all expert weights addressable.
Because routing follows attention, the selected experts are input-dependent and become known late.
Thus, each request combines a growing, mutable KV working set with a large, read-mostly expert-weight set that is accessed sparsely and late.

\subsubsection{Continuous Batching Serving}
These per-request memory characteristics become more dynamic in online serving, where requests with different prompt and generation lengths arrive asynchronously.
To maintain GPU utilization, modern serving systems commonly use continuous batching, admitting and removing requests at token boundaries rather than waiting for a fixed batch to complete~\cite{yu2022orca,sarathi_serve}.
As requests enter and leave, the active batch, prefill/decode mixture, aggregate KV footprint, and union of selected experts continuously change~\cite{sarathi_serve,yu2022orca,shazeer2017moe}.
The memory system must therefore manage large expert weights and growing mutable KV state under time-varying demand.

\subsection{High-Bandwidth Flash}
\label{sec:bg-hbf}

\paragraph{HBF Structure}
High-Bandwidth Flash (HBF) is a stacked memory designed to combine the high storage density of NAND flash with high aggregate read bandwidth.
HBF stack vertically integrates multiple flash core dies above a logic base die, with the dies interconnected by through-silicon vias (TSVs).
Because each die contains multiple NAND planes, page reads can proceed in parallel across dies and planes~\cite{sandisk_hbf,sandisk_hbf_patent_2025}.
This structure can provide hundreds of gigabytes of capacity (e.g., 512\,GB) and a reported stack-level read bandwidth of 1.6\,TB/s~\cite{sandisk_hbf}, making HBF attractive for storing large model weights and KV cache.

\paragraph{HBF Characteristics}

HBF retains NAND semantics: reads and programs operate on pages, erasure occurs at block granularity, and updates require out-of-place writes followed by garbage collection~\cite{kioxia_tc58_2019,agrawal2008ssd}. Each HBF read incurs NAND sensing latency $t_R$ before data transfer begins (e.g., 1--30\,$\mu$s), favoring large parallel transfers over fine-grained random accesses~\cite{kioxia_tc58_2019}.

Predictable accesses can hide $t_R$ through prefetching and double buffering.
In contrast, it may become exposed when the target address is determined late or when requests are too small to exploit parallelism across dies and planes.
HBF programming has substantially higher latency and lower bandwidth than reading, with $t_{\mathrm{PROG}}$ typically ranging from tens to hundreds of microseconds (e.g., 100\,$\mu$s), so fine-grained writes can stall inference~\cite{kioxia_tc58_2019}.

Finally, finite program/erase endurance makes sustained writes a lifetime concern~\cite{hong2022guardederase}.
Read-only model weights cause little write wear, whereas frequently updated KV cache requires efficient write scheduling and endurance management.
\section{Motivation}
\label{sec:motivation}

\subsection{Limitations of Existing Capacity Scaling}
\label{sec:motivation-capacity}
\begingroup
\renewcommand{\footnotesize}{\fontsize{9pt}{10.5pt}\selectfont}

\begin{table}[t]
  \centering
  \caption{comparison of memory-capacity expansion approaches}
  \label{table:motivation_table}
  \vspace{-1.5pt}

  \fontsize{9pt}{9pt}\selectfont
  \renewcommand{\arraystretch}{1.0}
  \setlength{\tabcolsep}{1.0pt}
  \setlength{\doublerulesep}{1.2pt}

  \newcommand{\tableheadercell}[2]{%
    \parbox[c][19pt][c]{#1}{%
      \centering\fontsize{9pt}{9pt}\selectfont #2%
    }%
  }
  \newcommand{\tablebodycell}[2]{%
    \parbox[c][21pt][c]{#1}{%
      \centering\fontsize{9pt}{9pt}\selectfont #2%
    }%
  }
  \newcommand{\approachcell}[1]{%
    \tablebodycell{0.16\columnwidth}{#1}%
  }
  \newcommand{\benefitcell}[1]{%
    \tablebodycell{0.18\columnwidth}{#1}%
  }
  \newcommand{\issuecell}[1]{%
    \tablebodycell{0.25\columnwidth}{#1}%
  }
  \newcommand{\limitcell}[1]{%
    \tablebodycell{0.22\columnwidth}{#1}%
  }
  \newcommand{\costcell}[1]{%
    \tablebodycell{0.14\columnwidth}{#1}%
  }

  \begin{tabular}{@{}ccccc@{}}
    \hline

    \tableheadercell{0.16\columnwidth}{\textbf{Approach}}
    & \tableheadercell{0.18\columnwidth}{%
        \textbf{Capacity /}\\\textbf{bandwidth}}
    & \tableheadercell{0.25\columnwidth}{%
        \textbf{Decode-path}\\\textbf{issue}}
    & \tableheadercell{0.22\columnwidth}{%
        \textbf{Main}\\\textbf{limitation}}
    & \tableheadercell{0.14\columnwidth}{\textbf{Cost}} \\ \hline

    \approachcell{Multi-GPU}
    & \benefitcell{Yes / High}
    & \issuecell{GPU traffic\\overhead}
    & \limitcell{Extra GPUs}
    & \costcell{High} \\ \hline

    \approachcell{CPU\\offload}
    & \benefitcell{Yes / Low}
    & \issuecell{PCIe transfer\\bottleneck}
    & \limitcell{Host-memory\\stalls}
    & \costcell{Low to\\medium} \\ \hline

    \approachcell{SSD offload}
    & \benefitcell{Yes / Low}
    & \issuecell{NVMe access\\latency}
    & \limitcell{On-demand\\read stalls}
    & \costcell{Low} \\ \hline

    \rowcolor{black!10}
    \approachcell{HBF}
    & \benefitcell{Yes / Moderate}
    & \issuecell{Read/program\\latency}
    & \limitcell{Write cost /\\endurance}
    & \costcell{Not\\reported} \\ \hline
  \end{tabular}
  \vspace{-3pt}
\end{table}

\endgroup

As shown in Table~\ref{table:motivation_table},
CPU-based and SSD-based offloading schemes place model weights or KV cache blocks in host memory and storage respectively.
Tightly coupled CPU--GPU systems such as Grace Hopper substantially increase GPU--host-memory bandwidth through NVLink-C2C~\cite{nvlink_c2c}, but the bandwidth remains well below the GPU's local HBM bandwidth.

Multi-GPU model parallelism expands memory capacity by aggregating GPU memory, but necessarily scales compute capacity with it.
Although Qwen3-235B-A22B~\cite{qwen3_235b} activates only 22B parameters per token, storing its full 235B-parameter BF16 weight set requires approximately 470\,GB, nearly exhausting the nominal 480\,GB aggregate HBM2e capacity of six 80-GB NVIDIA H100 PCIe GPUs~\cite{h100_pcie} before reserving memory for the KV cache, activations, communication buffers, and runtime workspace.
Because autoregressive decoding is often memory-bandwidth bound, GPUs added primarily to satisfy memory-capacity requirements can leave much of their compute underutilized~\cite{pope2023scaling,sarathi_serve}.
Tensor parallelism incurs communication and synchronization overhead from inter-GPU collectives, whereas pipeline parallelism adds inter-stage transfers and pipeline bubbles~\cite{pope2023scaling,narayanan2021megatron}.
In MoE models, input-dependent routing can further create load imbalance by concentrating tokens on experts hosted by particular GPUs~\cite{hwang2023tutel}.
Thus, scaling out H100 GPUs merely to fit model weights increases system overhead even when the added compute is not fully utilized.

\subsection{Latency Challenges in HBF}
\label{sec:motivation-challenges}
Direct GPU access alone is insufficient to use HBM and HBF efficiently.
Data placement must balance traffic across both memories, expose die- and plane-level HBF parallelism, and separate long-lived model weights from frequently updated KV cache.
Each HBF read incurs the NAND sensing latency, $t_R$, before data transfer begins.
Predictable dense-layer accesses can hide $t_R$ through prefetching, whereas MoE expert accesses depend on routing results.
If an expert read is issued only after routing completes, $t_R$ is exposed on the inference critical path.
Expert selection must therefore begin early enough to overlap $t_R$ with preceding computation. HBF writes incur the substantially longer page-program latency, $t_{\mathrm{PROG}}$.
Prefill produces large KV cache bursts that can be combined into page-sized writes, whereas decode generates small per-token updates that poorly utilize the program granularity.
Issuing these writes immediately would repeatedly expose $t_{\mathrm{PROG}}$ and interfere with HBF reads.

Our design addresses these challenges using independent paths, HBF-aware data placement, early expert selection to hide $t_R$, and phase-aware KV buffering and scheduling to mitigate the impact of $t_{\mathrm{PROG}}$.
\section{\systemname{} Architecture}
\label{sec:architecture}

\begin{figure}[!t]
  \centering
  \includegraphics[width=\linewidth]{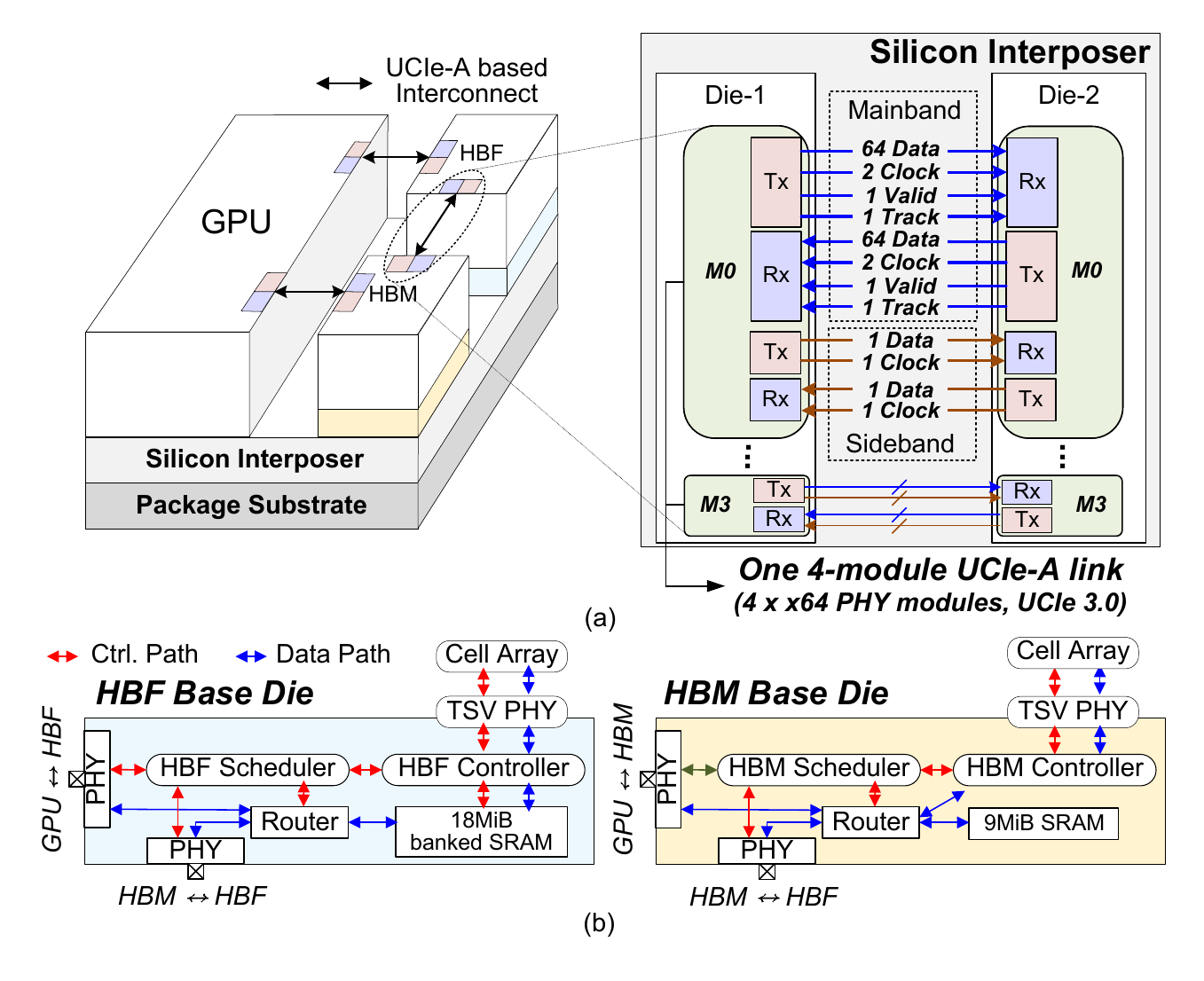}
  \caption{(a) DASH architecture and a four-module UCIe~3.0
UCIe-A link provisioned for 1.6~TB/s of usable D2D
bandwidth; (b) microarchitectures of the HBF and HBM base
dies.}
  \label{fig:fig_4}
\end{figure}

We propose \systemname{}, a heterogeneous near-GPU memory architecture that integrates HBM and HBF.
Figure~\ref{fig:fig_4}(a) shows the overall architecture of \systemname{}, where HBM and HBF operate as independent near-GPU memories.
\systemname{} provides three data paths: GPU--HBM, GPU--HBF, and HBM--HBF.
These paths are implemented using Universal Chiplet Interconnect Express (UCIe) links that connect the GPU I/O die to the HBM and HBF base dies and directly connect the two memory base dies~\cite{ucie_spec3,das_sharma2025ucie_memory}.

\subsection{UCIe-Based Unified Memory Interconnect}
\label{sec:ucie_interconnect}
\systemname{} equips the GPU I/O die and both memory base dies with UCIe 3.0 physical layers (PHYs) and die-to-die (D2D) adapters, providing a common transport across the three data paths~\cite{ucie_spec3,sharma2022universal,ucie_sip_2024}.
A common transaction format carries the operation, address, transfer length, and data over the UCIe mainband, while the receiving scheduler routes each request to the appropriate memory controller, which translates it into HBM- or NAND-specific commands~\cite{ucie_spec3,sharma_coughlin2024ucie_memory_storage,das_sharma2025ucie_memory}.

Each path is implemented as one four-module UCIe-A link, whose common adapter and multi-module PHY logic coordinate four paired $\times$64 modules at 64\,GT/s as a single logical link~\cite{ucie_spec3,sharma2022universal,ucie_sip_2024}.
The resulting 2.048\,TB/s raw capacity per direction leaves approximately 22\% headroom for link and implementation losses while supporting 1.6\,TB/s of modeled usable D2D streaming bandwidth per physical link~\cite{ucie_spec3,sharma2022universal}.
Each module retains an independent sideband for link training, configuration, and repair, whereas the common adapter presents one transaction stream to the base-die router~\cite{ucie_spec3}.
NAND sensing and shared-TSV SRAM fills are accounted for separately from this D2D rate and overlapped through double buffering.

Building on UCIe-based attachment of memory base dies~\cite{sharma_coughlin2024ucie_memory_storage,das_sharma2025ucie_memory}, \systemname{} relays HBF reads through the HBM base-die router to the GPU-facing link without accessing the HBM controller or cell array.
We refer to this end-to-end HBF--HBM-base-die--GPU route as the \emph{Relay path}.
UCIe 3.0 defines UCIe-A energy-efficiency targets of approximately 0.5\,pJ/bit at 0.5\,V and 0.6\,pJ/bit at 0.7\,V for the relevant high-speed operating points, and specifies a 0.389-mm-wide UCIe-A module~\cite{ucie_spec3}.
Four such modules therefore require roughly 1.6\,mm of die-edge width per endpoint~\cite{ucie_spec3,ucie_sip_2024}.
At 1.6\,TB/s, these energy-efficiency targets correspond to approximately 6.4--7.7\,W of link power per fully utilized direction~\cite{ucie_spec3}.
These specification-based estimates characterize the die-edge footprint and link power, while physical design further accounts for routing, signal- and power-integrity, and power-delivery costs~\cite{sharma2022universal,ucie_sip_2024,das_sharma2025ucie_memory,wu_guo2025ucie64g}.

\subsection{HBM and HBF Base-Die Microarchitecture}
\label{sec:base-die}

Figure~\ref{fig:fig_4}(b) illustrates the base-die microarchitecture of HBM and HBF of \systemname{} architecture.

\paragraph{HBF Base Die}
The HBF base die handles read and write requests received from the GPU or the HBM side through UCIe.
It contains an HBF controller, a local scheduler, and 18\,MiB of physical banked SRAM per stack.
Of this capacity, 16\,MiB is usable data storage organized into two interleaved 8\,MiB transfer regions.
The additional 2\,MiB accounts for 12.5\% ECC check-bit storage, assuming eight check bits per 64 data bits.
Each transfer region matches one stack-local page wave in our evaluation.
The two transfer regions alternate between shared-TSV fill and D2D drain.
Each read still incurs $t_R$ before its data become ready in SRAM.
Independently accessible banks allow different ready chunks to drain concurrently over the GPU--HBF path and \emph{Relay path}.
The scheduler issues a request only when the target die or plane, sufficient SRAM space, and the required transfer path are available.
The HBF controller then converts the request into NAND read or program commands.
Incoming writes use an available transfer region and are backpressured when both regions are occupied.

\paragraph{HBM Base Die}
The HBM-side relay stream is segmented into 4\,MiB transfer chunks, each equal to half of the 8-MiB stack-local HBF page wave.
Each HBM base die provisions 9\,MiB of physical banked SRAM per stack.
Of this capacity, 8\,MiB is usable relay storage organized as two interleaved 4-MiB transfer regions.
The additional 1\,MiB accounts for the same 12.5\% ECC check-bit storage.
The two transfer regions alternate between receiving one chunk from HBF and forwarding the preceding chunk to the GPU.
Because the HBF--HBM and GPU--HBM links have the same 1.6\,TB/s rate, this organization sustains the \emph{Relay path} after the initial fill.
When both regions are occupied, the scheduler withholds credits for \emph{Relay path} until one region becomes available.
The \emph{Relay path} data bypasses the HBM controller and cell array, so this transfer does not access DRAM.
Small buffer metadata and request and credit queues are provisioned separately and are not included in the quoted SRAM capacities.

Overall, \systemname{} combines independent GPU access to HBM and HBF with direct data movement between the two memories.
This architecture allows HBF data to use the GPU--HBF path alone or additionally use the GPU--HBM path through the \emph{Relay path}.
\section{Placement and Execution}
\label{sec:placement-execution}

\subsection{Data Placement}
\label{sec:data-placement}
\subsubsection{Placement Along Architecture}
\systemname{} places data according to size, mutability, and reuse: HBM holds frequently updated state, HBF holds large read-mostly data, and small frequently reused read-only weights may be replicated.
Expert weights and write-once/read-many prefill KV reside in HBF, while smaller attention weights are replicated across HBM and HBF.
Fine-grained decode KV is initially accumulated in HBM and migrated to HBF in complete page-aligned units before HBM fills.
\begin{figure}[t]
    \centering
    \includegraphics[width=\linewidth]{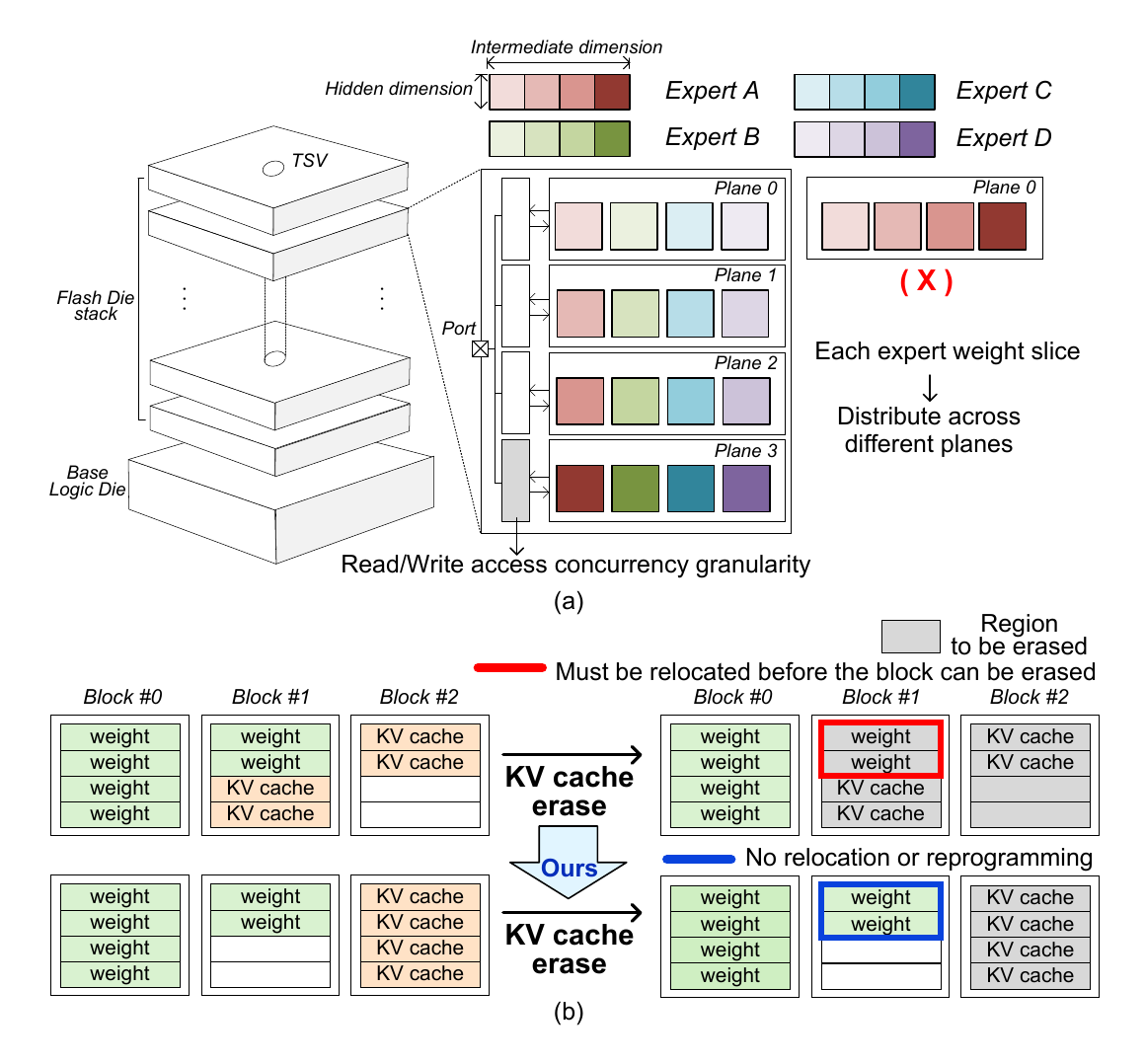}
    \caption{(a) Placement of MoE expert weights across HBF planes for parallel access
    (b) KV cache placement with erase block granularity}
    \label{fig:fig_5}
\end{figure}
\subsubsection{Placement Within HBF}
For MoE weights, mapping each expert to a single plane or a small subset of planes would limit the available read bandwidth.
\systemname{} therefore divides each expert's weights into chunks and distributes them across multiple HBF dies and planes.
As shown in Figure~\ref{fig:fig_5}(a), this placement allows the weights of a selected expert to be read in parallel and utilize high HBF read bandwidth even when only a few experts are activated.

Because HBF erases at block granularity, garbage collection may relocate live pages~\cite{agrawal2008ssd}.
To isolate model weights from KV garbage collection, as illustrated in Figure~\ref{fig:fig_5}(b), \systemname{} stores model weights and KV cache in separate weight-owned and KV-owned blocks.
During garbage collection, a KV-owned block is erased directly if it contains no live KV pages.
This separation allows KV garbage collection to proceed without examining or relocating model weight pages, reducing data-relocation traffic.

\begin{figure*}[!t]
    \centering
    \includegraphics[width=\linewidth]{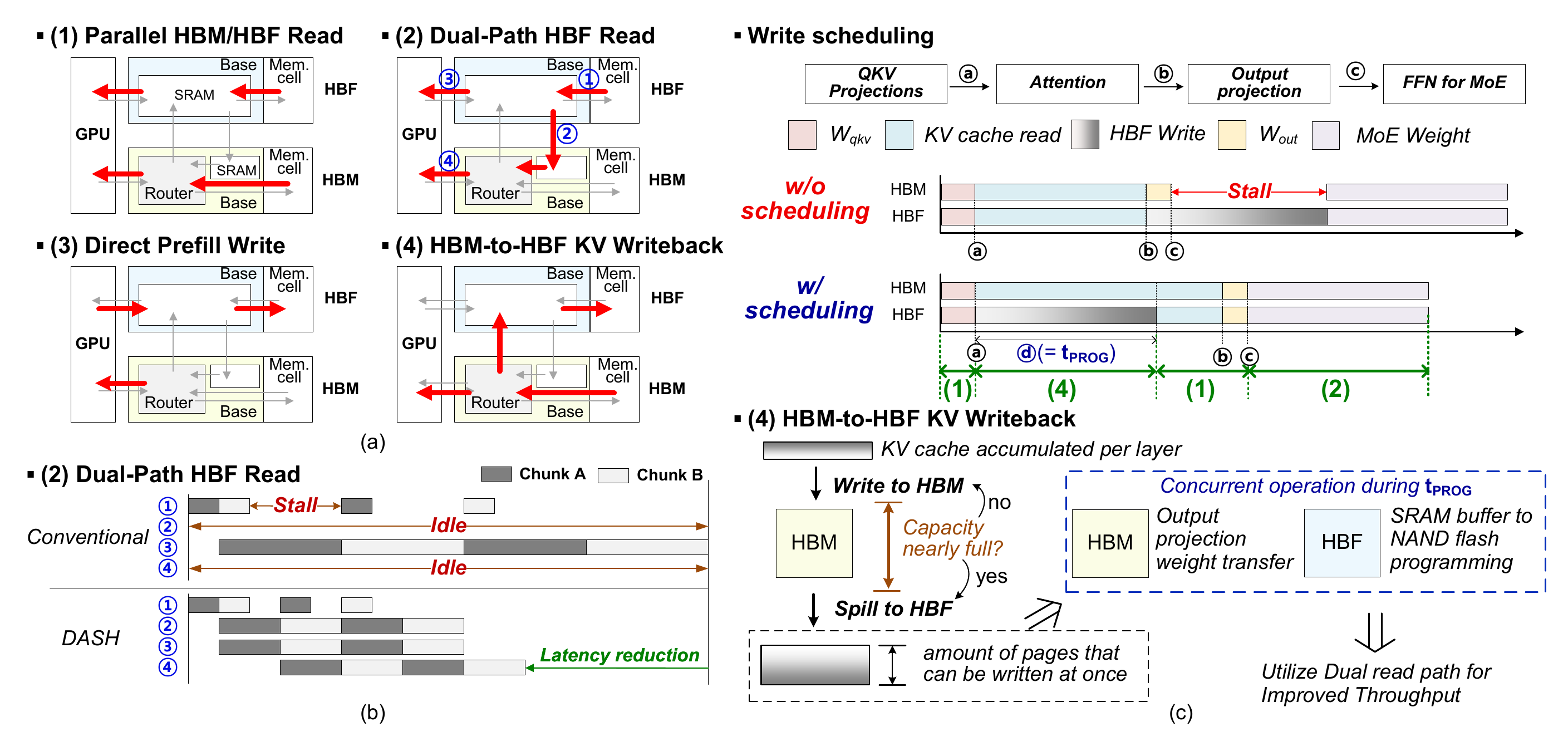}
    \caption{(a) Four primary execution modes of \systemname{};
    (b) latency reduction with dual-path HBF reads;
    (c) effect of HBF write scheduling. }
    \label{fig:fig_6}
    \vspace{-5pt}
\end{figure*}

\subsection{Dataflow and Execution}
\label{sec:dataflow-execution}
The GPU-side scheduler routes each request to the memory holding its data through an available path with buffer space.
Figure~\ref{fig:fig_6}(a) summarizes four representative execution modes.

\paragraph{Parallel HBM/HBF Read}
In this mode, the GPU concurrently reads distinct data from HBM and HBF through their respective paths.
First, different portions of the replicated attention weights can be read from HBM and HBF in parallel, increasing the aggregate bandwidth available to the GPU.
Second, attention can simultaneously read the KV cache ranges placed in HBM and HBF.
The GPU computes partial attention results for each range and combines them before producing the final attention output.

\paragraph{Dual-Path HBF Read}
The dual-path HBF read uses independent SRAM banks to concurrently deliver different HBF-resident chunks over both GPU-facing paths.
NAND reads still incur $t_R$, and each chunk becomes transferable only after it is filled into an SRAM bank through the shared TSV.
Double buffering overlaps this preparation with transmission of previously ready chunks.
Direct reads move from HBF-side SRAM to the GPU, while relayed reads traverse HBM-side SRAM and the GPU--HBM path without accessing the HBM cell array.
Separate SRAM banks allow both paths to operate concurrently once their data are ready.

\systemname{} uses this mode when additional delivery bandwidth is needed for selected expert weights or HBF-resident KV cache.
The conventional design leaves one GPU-facing path idle while the other drains ready data, resulting in underutilized bandwidth and a longer completion time.
As shown in Figure~\ref{fig:fig_6}(b), \systemname{} simultaneously drains different ready chunks over both GPU-facing paths to increase aggregate delivery bandwidth and reduce transfer latency.

\paragraph{Direct Prefill Write}
During prefill, \systemname{} writes newly generated KV cache directly from the GPU to HBF.
Because prefill generates a large volume of KV cache at once, its writes achieve high utilization of the HBF program granularity.
The generated KV cache is sent through the GPU--HBF path to the SRAM of HBF and is then programmed into HBF by the HBF controller.
Direct prefill write execution sends the persistent KV copy from the GPU to HBF without staging it in HBM.
Because the GPU--HBM and GPU--HBF paths operate independently, the current layer's output-projection weights can be fetched from HBM while HBF programming overlaps the attention computation.
The compute-intensive attention operation provides a long interval in which the HBF program latency can be hidden. 
This scheduling avoids serializing the prefill KV write with the subsequent attention computation.

\paragraph{HBM-to-HBF KV Writeback}
Since HBF program latency is substantially longer than read latency, write timing is crucial.
During decode, \systemname{} appends small KV entries to HBM because individual entries are often too fine-grained for efficient HBF programming~\cite{kioxia_tc58_2019,agrawal2008ssd}.
A full HBF page wave is the aggregate data volume that can be programmed concurrently across all HBF stacks, dies, and planes.
Once HBM approaches its capacity limit and each HBM stack has accumulated its partition of a full page wave, \systemname{} writes the wave back in parallel over the HBM--HBF paths, as shown in Figure~\ref{fig:fig_6}(c).
The transfer begins after the current layer finishes its QKV projection, allowing the HBF program latency to overlap with the following attention computation.
The attention computation uses the KV cache in HBM, while older KV entries are transferred to HBF through the HBM--HBF path.
The HBM controller transfers the selected KV data to the HBF-side SRAM over the HBM–HBF path, and the HBF controller then writes it into HBF.
After writeback, \systemname{} records the KV data’s new HBF location and can free the corresponding HBM space.
Batching small decode updates into larger HBF writes avoids a separate programming delay on every decode iteration.

\subsection{Lookahead Expert Execution}

Although HBF provides high read bandwidth, each read incurs the NAND sensing latency $t_R$~\cite{sandisk_hbf,sandisk_hbf_patent_2025,kioxia_tc58_2019}.
For predictable access sequences, \systemname{} hides this latency using double buffering in the HBF base die.
While one HBF-side SRAM buffer supplies the current data to the GPU, the other prepares data for the next access.
For non-expert weights and scheduled KV cache accesses, upcoming addresses follow the model execution order and can be identified in advance.

However, this approach does not directly apply to MoE expert accesses~\cite{huang2024moeinference,zhu2025preattention}.
Because MoE routing is input-dependent, the selected experts are not known until routing completes~\cite{shazeer2017moe,fedus2022switch,jiang2024mixtral,deepseekv2,deepseekv3}.
In conventional execution, the attention block is followed by the output projection, residual addition, and RMSNorm, after which routing begins~\cite{vaswani2017attention,rmsnorm2019,jiang2024mixtral,deepseekv2,deepseekv3}.
When expert weights reside in HBF, their reads cannot be issued until this point, exposing the initial HBF read latency~\cite{sandisk_hbf_patent_2025,h3,hbm_hbf_pooling,huang2024moeinference}.
Prior work~\cite{hbm_hbf_pooling} avoids this latency by keeping MoE expert weights in HBM.
However, the expert weights of large-scale MoE models can exceed the available HBM capacity, making this approach impractical~\cite{mixtral8x22b,qwen3_235b,llama4_maverick,deepseekv3,b200}.
Therefore, when expert weights are placed in HBF, the selected experts must be determined early enough to initiate their reads~\cite{zhu2025preattention,madan2026speculating}.

\begin{figure}[t]
    \centering
    \includegraphics[width=\linewidth]{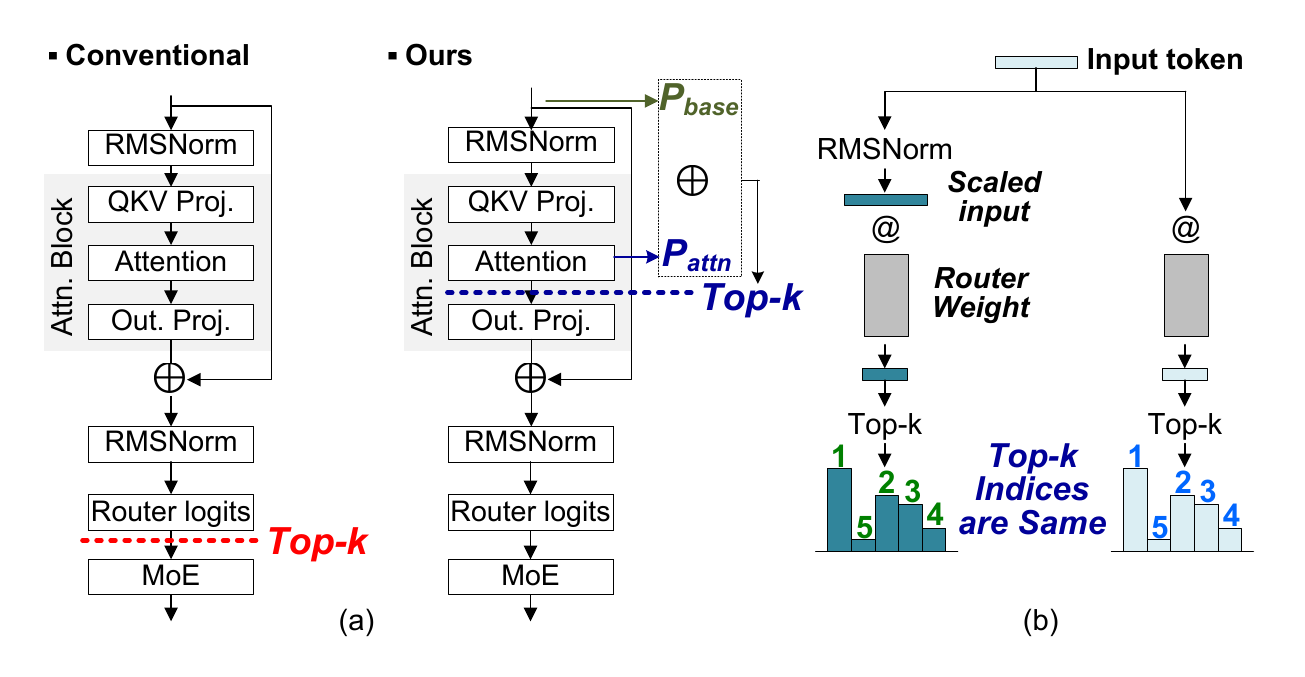}
    \caption{(a) Comparison of MoE top-k selection timing (b) Equivalent top-$k$ selection without waiting for RMS scaling}
    \label{fig:fig_7}
\end{figure}

\systemname{} decomposes conventional routing into an input-only term computed before attention and an output-dependent term computed as soon as the attention output becomes available.
Combining these terms determines the top-$k$ experts before conventional routing, as shown in Figure~\ref{fig:fig_7}(a).

Let $X$ denote the input to the RMSNorm preceding the attention block, $O$ the attention output before the output projection, $W_{\mathrm{out}}$ the output projection weight, $\gamma$ the scale parameter of the RMSNorm preceding the MoE block, and $W_{\mathrm{r}}$ the routing weight.
The conventional routing logits are
\begin{equation}
P
=
\operatorname{RMSNorm}
\left(
X + OW_{\mathrm{out}}
\right)
W_{\mathrm{r}}.
\end{equation}
Using the definition of RMSNorm~\cite{rmsnorm2019}, these logits can be rewritten as
\begin{equation}
P
=
\alpha
\left(
P_{\mathrm{base}} + P_{\mathrm{attn}}
\right),
\end{equation}
where
\begin{equation}
P_{\mathrm{base}}
=
X\operatorname{diag}(\gamma)W_{\mathrm{r}},
\end{equation}
\begin{equation}
P_{\mathrm{attn}}
=
O\widehat{W}_{\mathrm{r}},
\qquad
\widehat{W}_{\mathrm{r}}
=
W_{\mathrm{out}}\operatorname{diag}(\gamma)W_{\mathrm{r}},
\end{equation}
and
\begin{equation}
\alpha
=
\frac{1}{
\operatorname{RMS}
\left(
X + OW_{\mathrm{out}}
\right)
}.
\end{equation}

The base-routing term, $P_{\mathrm{base}}$, depends only on $X$ and can therefore be computed before the attention computation begins.
The attention-routing term, $P_{\mathrm{attn}}$, can be computed as soon as $O$ becomes available, without waiting for completion of output projection.
Because $\widehat{W}_{\mathrm{r}}=W_{\mathrm{out}}\operatorname{diag}(\gamma)W_{\mathrm{r}}$ is constant during inference, it can be precomputed and stored as an additional weight.
Thus, \systemname{} obtains $P_{\mathrm{base}}+P_{\mathrm{attn}}$ before the conventional routing point.

The scale factor $\alpha$ is a positive scalar shared by all expert logits of the same token.
It changes their magnitudes but does not affect the rank.
Therefore, selecting the top-$k$ experts from $P_{\mathrm{base}}+P_{\mathrm{attn}}$ produces the same routing decision as conventional routing, as shown in Figure~\ref{fig:fig_7}(b).

\systemname{} therefore determines the top-$k$ experts without waiting for the RMS value.
The early routing terms are used only for expert selection.
Once the RMSNorm output becomes available, the subsequent FFN operation can be done with the expert weight selected beforehand.
This allows the HBF controller to begin reading the selected expert weights as soon as the attention-routing term completes~\cite{zhu2025preattention,madan2026speculating}.
Conventional execution instead waits for the residual addition, RMSNorm, and routing computation, leaving HBF idle between the output projection and expert-weight streams~\cite{jiang2024mixtral,deepseekv2,deepseekv3}.
\systemname{} minimizes this read stall and extends HBF latency hiding to input-dependent MoE expert accesses.

\systemname{} computes expert selection in FP32 to ensure accurate selection.
It uses early selection only when routing is scale-invariant and free of expert-specific additive bias.
For all other routers, including the router used in DeepSeek-V3, \systemname{} performs conventional late selection while preserving its placement and dual-path delivery~\cite{deepseekv3}.
All other model computations remain at their native precision.
\section{Evaluation}
\label{sec:evaluation}

\subsection{Methodology}
\label{sec:eval-method}

\paragraph{Models}
\begingroup
\renewcommand{\footnotesize}{\fontsize{9pt}{10.5pt}\selectfont}

\begin{table}[t]
  \centering
  \caption{Evaluated MoE models and configurations}
  \label{tab:eval-models}

  \fontsize{9pt}{9.4pt}\selectfont
  \setlength{\tabcolsep}{1.2pt}
  \renewcommand{\arraystretch}{1.03}
  \renewcommand{\tabularxcolumn}[1]{m{#1}}

  \newcommand{\EvalHeader}[1]{%
    \parbox[c][21pt][c]{\linewidth}{%
      \centering\fontsize{9pt}{9pt}\selectfont #1%
    }%
  }
  \newcommand{\EvalModelHeader}[1]{%
    \parbox[c][21pt][c]{\linewidth}{%
      \raggedright\fontsize{9pt}{9pt}\selectfont #1%
    }%
  }

  \begin{tabularx}{0.98\columnwidth}{@{}
    >{\raggedright\arraybackslash}X
    >{\centering\arraybackslash}m{0.10\columnwidth}
    >{\centering\arraybackslash}m{0.16\columnwidth}
    >{\centering\arraybackslash}m{0.19\columnwidth}
    >{\centering\arraybackslash}m{0.20\columnwidth}@{}}
    \hline

    \EvalModelHeader{Model}
    & \EvalHeader{Attn.}
    & \EvalHeader{Weights\\(GB)}
    & \EvalHeader{Experts /\\Top-$k$}
    & \EvalHeader{Weight Precision} \\

    \hline
    \noalign{\vskip 1.5pt}
    
    Qwen3-235B-A22B
    & GQA & 470.19 & 128+0 / 8 & BF16 \\

    Mixtral-8$\times$22B
    & GQA & 281.24 & 8+0 / 2 & BF16 \\

    Grok-1
    & GQA & 318.24 & 8+0 / 2 & INT8-mixed \\

    Llama~4 Maverick
    & GQA & 801.42 & 128+1 / 1 & BF16 \\

    DeepSeek-V3
    & MLA & 688.59 & 256+1 / 8 & FP8-mixed \\

    DeepSeek-V2
    & MLA & 471.49 & 160+2 / 6 & BF16 \\

    \hline
  \end{tabularx}
\end{table}

\endgroup

\begin{figure}[!t]
  \centering
  \includegraphics[width=\linewidth]{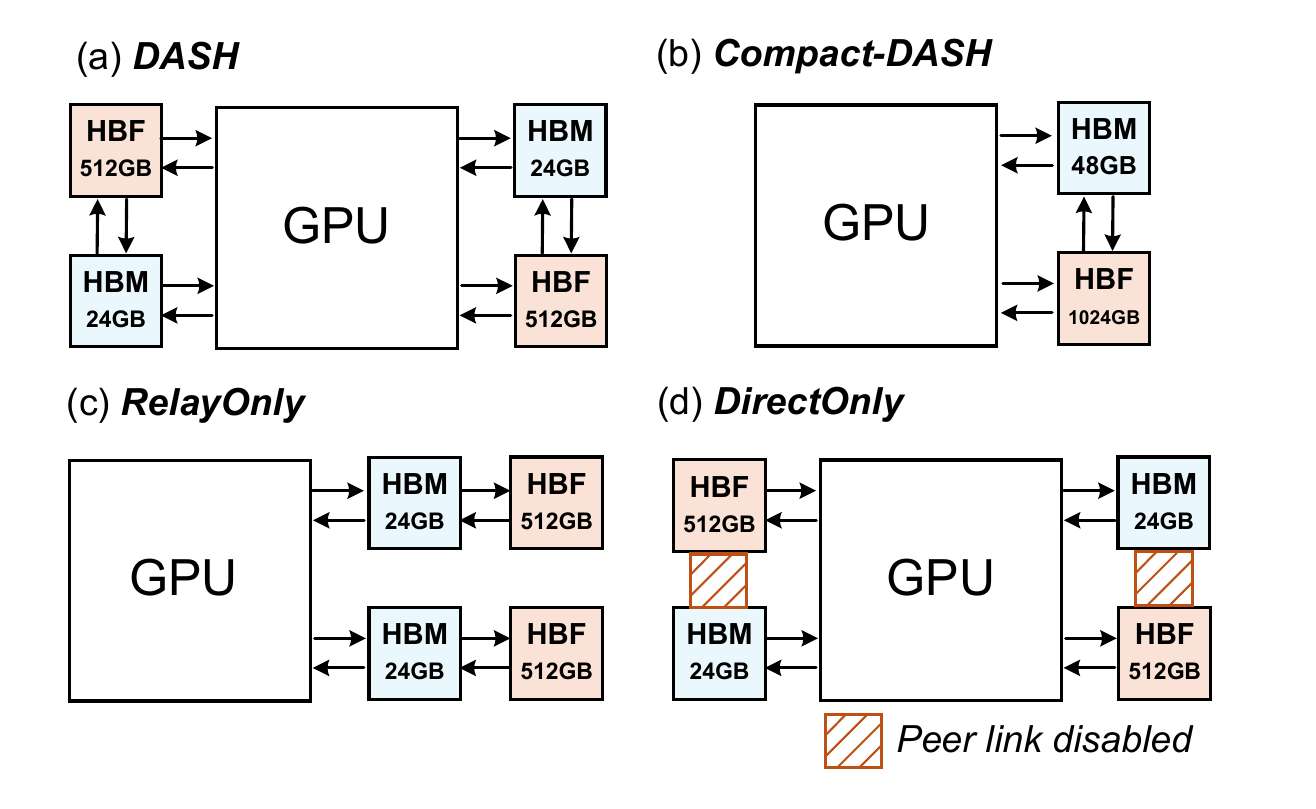}
  \caption{Four configurations used in the evaluation.}
  \label{fig:fig_8}
  \vspace{-2pt}
\end{figure}
\begin{figure*}[!t]
  \centering
  \includegraphics[width=0.99\textwidth]{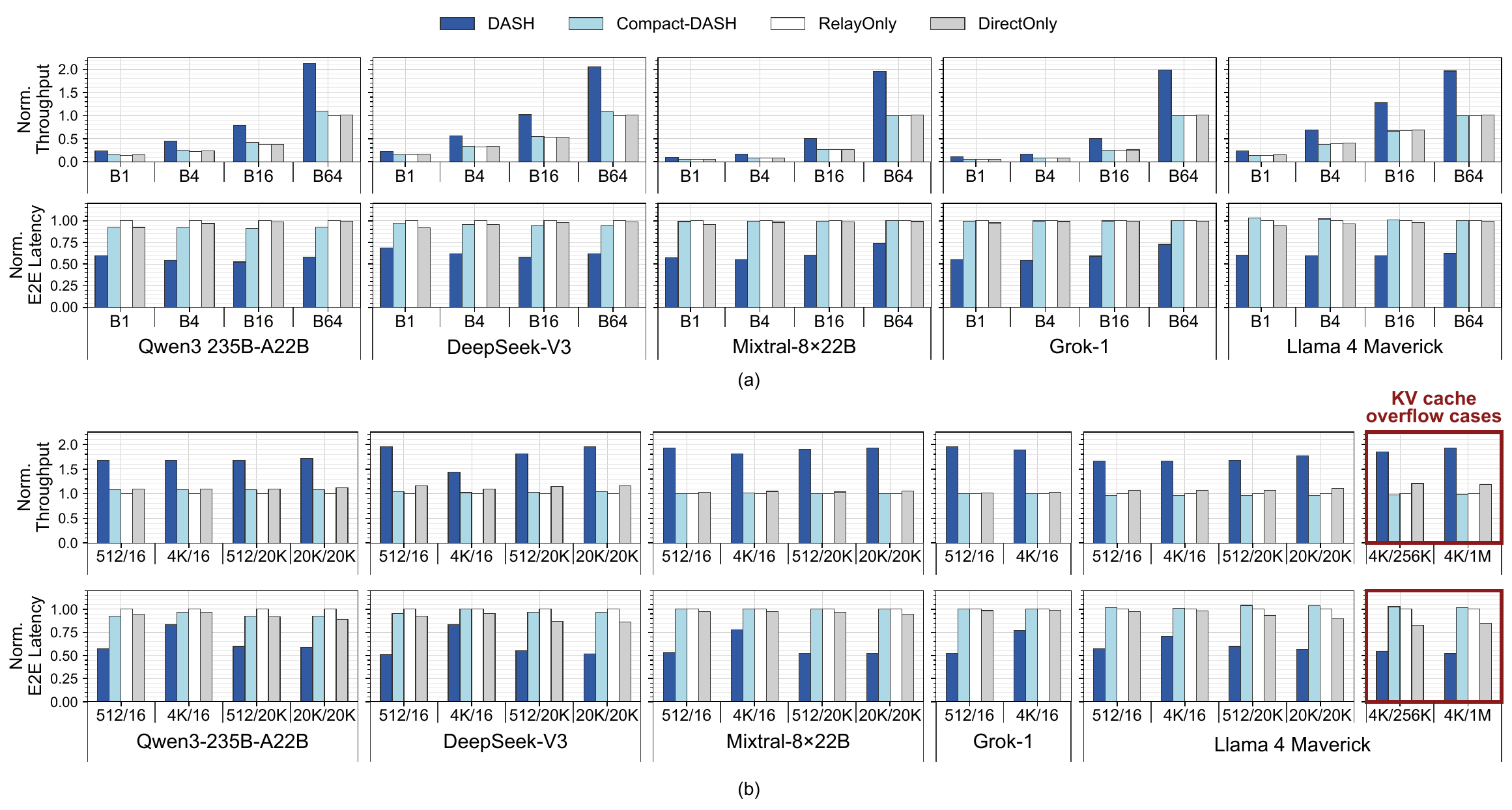}
  \caption{Normalized throughput and E2E latency under (a) batch-size scaling and (b) B1 input/output sequence-length variation.
  In (b), each x-axis label reports $L_{\mathrm{in}}/L_{\mathrm{out}}$, the input/output sequence lengths.
  Red boxes mark the Llama 4 Maverick KV cache overflow cases.}
  \label{fig:fig_9}
\end{figure*}

Table~\ref{tab:eval-models} summarizes six MoE model configurations~\cite{qwen3_235b,mixtral8x22b,grok1,llama4_maverick,deepseekv3,deepseekv2}. 
Five are evaluated in the main sweeps over batch size and sequence length, while DeepSeek-V2 is used separately for the Lookahead Expert Execution study described in Section~\ref{sec:eval-eed}. The Experts column lists the numbers of routed and shared experts.

\paragraph{Baselines and HBF Configuration}
Figure~\ref{fig:fig_8} illustrates four architectural configurations used to evaluate the effectiveness of \systemname{}.
Across all configurations, we use the consistent data-placement policy: expert weights reside in HBF, QKV and output projection weights are replicated across HBM and HBF.
The \systemname{} configuration consists of two 512\,GB HBF stacks and two 24\,GB HBM stacks~\cite{micron_hbm3e,sandisk_hbf}, with each physical D2D link provisioned for 1.6\,TB/s of usable streaming bandwidth.
NAND/TSV preparation and SRAM readiness are modeled separately from this link rate.
RelayOnly and DirectOnly preserve the same number of memory stacks while removing the direct GPU--HBF path and the HBF--HBM path, respectively.
Compact-\systemname{} retains both paths but consolidates the memory resources into a single HBM/HBF stack pair, enabling comparison under a matched GPU-side connectivity budget with area reduction.

We model 4-KiB HBF pages following the physical-page organization described in the HBF patent~\cite{sandisk_hbf_patent_2025}.
Because public HBF disclosures do not specify the exact subarray-level parallelism, we assume four independently accessible subarrays per plane.
With 16 dies and 32 planes per die, this yields an 8\,MiB page wave per stack and 16 MiB across two stacks.
We use 3-$\mu$s read and 100-$\mu$s program latencies as nominal points and sweep these parameters in Section~\ref{sec:sensitivity}.

\paragraph{Simulation}

To evaluate \systemname{}, we extend LLMSimulator~\cite{llmsim} with HBF modeling and shape-aware GPU performance profiling.
The HBF model captures read and program latencies, die-level and plane-level parallelism, SRAM-buffer readiness, per-link D2D availability, and KV-cache writeback of adjacent paths.
We profile GPU operator latencies on an NVIDIA H100 PCIe 80\,GB GPU using the evaluated workloads~\cite{nvidia_h100_specs}.
The simulator directly uses measured latencies for profiled GPU operator shapes and interpolates unmeasured shapes only within validated per-operator ranges.
Across 1,107 validation measurements of supported non-router operators, the GPU timing model achieves a median relative error of 0.51\%, with 90\% of errors below 3.52\%.

We use $B$, $L_{\mathrm{in}}$, and $L_{\mathrm{out}}$ to denote batch size and input/output sequence lengths, respectively.
Unless otherwise stated, fixed-batch experiments use $B/L_{\mathrm{in}}/L_{\mathrm{out}} = 4/1\mathrm{K}/128$.

\subsection{Throughput and End-to-End Latency}
\label{sec:eval-overall}

Figure~\ref{fig:fig_9} compares the normalized throughput and E2E latency of \systemname{}, Compact-\systemname{}, RelayOnly, and DirectOnly.
Figure~\ref{fig:fig_9}(a) evaluates five MoE models at $B \in \{1,4,16,64\}$ with $L_{\mathrm{in}}/L_{\mathrm{out}}{=}1\mathrm{K}/128$.
Figure~\ref{fig:fig_9}(a) normalizes throughput to RelayOnly at $B{=}64$ for each model and E2E latency to RelayOnly at the corresponding batch size.
Figure~\ref{fig:fig_9}(b) fixes $B{=}1$ and varies the input/output sequence lengths, denoted by $L_{\mathrm{in}}/L_{\mathrm{out}}$: short $512/16$, prefill-dominant $4\mathrm{K}/16$, decode-heavy $512/20\mathrm{K}$, and long prefill+decode $20\mathrm{K}/20\mathrm{K}$.
Figure~\ref{fig:fig_9}(b) normalizes both metrics to RelayOnly for each model and workload.
Figure~\ref{fig:fig_9}(b) additionally evaluates Llama 4 Maverick KV cache overflow from HBM into HBF at $L_{\mathrm{in}}/L_{\mathrm{out}}{=}4\mathrm{K}/256\mathrm{K}$ and $4\mathrm{K}/1\mathrm{M}$.
We include only workloads that fit within each model's supported context window; consequently, Grok-1 is evaluated only at $512/16$ and $4\mathrm{K}/16$.

Figure~\ref{fig:fig_9}(a) shows that \systemname{} consistently outperforms RelayOnly and DirectOnly in both throughput and E2E latency at every batch size.
Across all five models and batch sizes, \systemname{} achieves geometric-mean throughput speedups of 1.90$\times$ over RelayOnly and 1.84$\times$ over DirectOnly, while reducing E2E-latency by 42.2\% and 40.8\%, respectively.
\systemname{} achieves these gains by transferring HBF-resident expert weights and KV cache over the Direct and Relay paths concurrently.
Despite halving the HBM/HBF stack-pair count, Compact-\systemname{} delivers performance comparable to RelayOnly and DirectOnly.
At each MoE layer, selected-expert transfers can use either the GPU--HBF Direct path or the HBM-mediated Relay path, enabling effective utilization of both transfer paths even within a single HBM/HBF pair.
Consequently, Compact-\systemname{} consistently outperforms the single-path layouts while providing a smaller, more area-efficient design point.

Figure~\ref{fig:fig_9}(b) shows that \systemname{} consistently delivers high throughput and low E2E latency across all evaluated workloads.
Across the 20 combinations, \systemname{} achieves geometric-mean throughput speedups of 1.79$\times$ and 1.63$\times$ over RelayOnly and DirectOnly, respectively, while reducing E2E latency by 40.1\% and 35.6\%, based on paired latency ratios.
In particular, compared with the prefill-dominated $4\mathrm{K}/16$ cases, the long-decode $512/20\mathrm{K}$ and $20\mathrm{K}/20\mathrm{K}$ cases exhibit larger E2E latency reductions while sustaining similarly high throughput across models.
Repeated decode steps create more opportunities for concurrent HBM--HBF access, while the proposed HBF write scheduling coordinates the growing KV cache traffic with latency-critical expert weight accesses.

The final two graphs evaluate Llama~4 Maverick with increasingly large KV caches.
For shorter decode workloads, the generated KV cache remains in HBM and does not trigger HBM-to-HBF writeback.
As the decode length increases, the KV cache grows beyond the available HBM capacity, causing older KV entries to be written back to HBF.
\systemname{} continues to achieve high throughput and low E2E latency even when HBM-to-HBF KV writeback occurs.
Even for Llama~4 Maverick with a 197.413\,GB KV cache generated by a long decode sequence, \systemname{} achieves 1.92$\times$ higher throughput and reduces E2E latency by 48.0\% over RelayOnly.
This robustness stems from \systemname{}'s two-level KV management: HBM absorbs fine-grained per-token updates, while batching them into larger writebacks through write scheduling reduces the HBF programming overhead per KV entry and reclaims HBM capacity.

\subsection{Comparison with CPU--GPU Offloading Strategies}
\label{sec:eval-cpu-hybrid-offload}
\begin{figure}[t]
  \centering
  \includegraphics[width=\columnwidth]
  {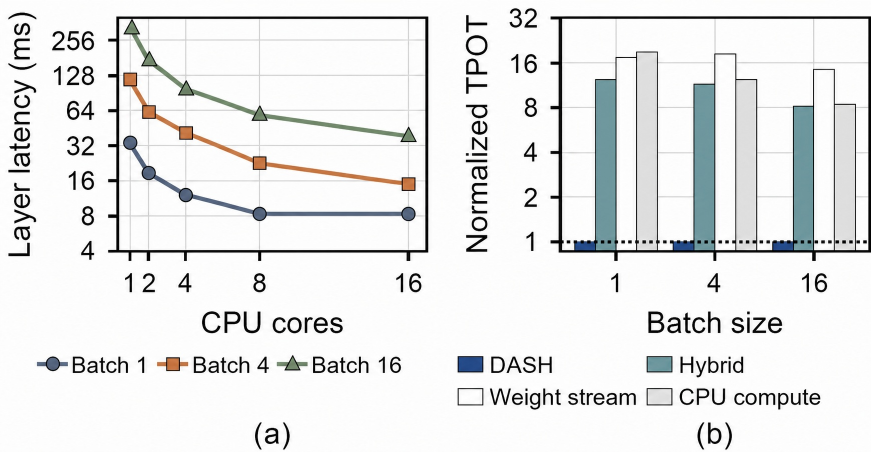}
  \caption{CPU-aware offloading:
  (a) median integrated Expert-layer latency under 1--16 CPU-core caps;
  (b) median-based TPOT normalized to DASH.
  CPU compute and endpoint-guarded Hybrid use the same 32-core cap in (b).}
  \label{fig:fig_10}
\end{figure}

When MoE weights exceed HBM, weight stream transfers each uncached expert's weights to the GPU~\cite{xue2024moeinfinity}, whereas CPU compute sends activations to CPU-resident experts~\cite{kamahori2025fiddler}.
We also evaluate a Hybrid oracle over all evaluated CPU/GPU workload splits, including the CPU-only and GPU-only configurations. For each split, we independently measure median TPOT; mixed splits execute CPU and GPU concurrently.
We report the minimum, providing an empirical upper bound on Hybrid performance, or equivalently, a lower bound on median TPOT, within the evaluated split space.

We measure a shape-equivalent Qwen3 BF16 expert layer on an Intel Xeon Platinum 8452Y in the H100-local NUMA node using physical cores~\cite{intel8452y}.
Figure~\ref{fig:fig_10}~(a) sweeps 1--16-core caps, while CPU compute and Hybrid in Figure~\ref{fig:fig_10}~(b) are independently tuned under the same 32-core cap.
This oracle-favorable comparison combines measured expert-layer costs with the same non-expert GPU time across all schemes.

Figure~\ref{fig:fig_10}~(a) shows that B1 saturates once its eight expert jobs are parallelized, whereas larger batches expose enough jobs to benefit from additional cores.
By adapting the CPU/GPU split, Hybrid oracle reduces TPOT by 2.8--29.4\% over the better single-path configuration across the evaluated batch sizes.
Even with this favorable baseline, Hybrid remains 8.22--12.32$\times$ slower than \systemname{} because \systemname{} avoids both CPU execution and host-PCIe weight staging.

\subsection{Impact of Lookahead Expert Execution}
\label{sec:eval-eed}
\begin{figure}[!t]
  \centering
  \includegraphics[width=\columnwidth]{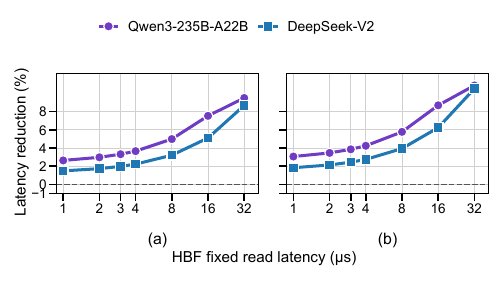}
  \caption{Impact of early expert decision on (a) request E2E latency and
  (b) time per output token.}
  \label{fig:fig_11}
\end{figure}
We isolate the effect of expert early decision within \systemname{} by comparing it against the conventional execution flow.
We measure the latency reduction at $B/L_{\mathrm{in}}/L_{\mathrm{out}}=1/4\mathrm{K}/128$.
Because concrete HBF specifications are not yet available, we further sweep the HBF read latency, $t_R$, from 1 to 32\,$\mu$s to evaluate expert early decision across a range of plausible HBF configurations.

As shown in Figure~\ref{fig:fig_11}, at 3\,$\mu$s, expert early decision reduces E2E latency by 3.33\% and 1.99\%, and TPOT by 3.86\% and 2.45\% for Qwen3 and DeepSeek-V2, respectively.
At 32\,$\mu$s, the corresponding E2E reductions increase to 9.50\% and 8.69\%, while the TPOT reductions reach 10.88\% and 10.53\%, respectively.
Lookahead Expert Execution exposes the target experts before conventional routing completes, allowing HBF read startup and weight delivery to overlap with the following output projection computation.
Thus, increasing $t_R$ initially exposes more latency that can be hidden from the expert-computation critical path.
Even when $t_R$ exceeds the interval between early selection and expert-weight consumption, the delay within this interval can only be overlapped; the excess remains on the critical path, but the overlapped portion still provides a clear reduction in both E2E latency and TPOT.

\begingroup
\renewcommand{\footnotesize}{\fontsize{9pt}{10.5pt}\selectfont}

\begin{table}[t]
  \centering
  \caption{Qwen3 P90 latency at 1.6\,TB/s per physical D2D LINK}
  \label{tab:table_3}
  \fontsize{9pt}{10pt}\selectfont
  \setlength{\tabcolsep}{1.2pt}
  \renewcommand{\arraystretch}{1.08}
  \renewcommand{\tabularxcolumn}[1]{m{#1}}
  \newcommand{\TableHeader}[1]{\parbox[c][25pt][c]{\linewidth}{\centering #1}}
  \begin{tabularx}{0.96\linewidth}{@{}
    >{\centering\arraybackslash}m{0.11\linewidth}
    >{\centering\arraybackslash}m{0.14\linewidth}
    >{\centering\arraybackslash}X
    >{\centering\arraybackslash}X
    >{\centering\arraybackslash}X@{}}
    \hline
    \TableHeader{Load}
    & \TableHeader{Metric}
    & \TableHeader{\systemname{}\\P90}
    & \TableHeader{Reduction\\vs. RelayOnly}
    & \TableHeader{Reduction\\vs. DirectOnly} \\
    \hline
    \multirow[c]{2}{*}{50\%}
      & E2E  & 20.982~s & 61.2\% & 61.0\% \\
      & TPOT & 148.0~ms & 62.2\% & 62.0\% \\
    \hline
    \multirow[c]{2}{*}{75\%}
      & E2E  & 28.847~s & 53.5\% & 53.3\% \\
      & TPOT & 208.5~ms & 53.9\% & 53.8\% \\
    \hline
    \multirow[c]{2}{*}{90\%}
      & E2E  & 31.859~s & 50.5\% & 50.4\% \\
      & TPOT & 232.0~ms & 50.3\% & 50.2\% \\
    \hline
  \end{tabularx}
\end{table}

\endgroup

\subsection{Continuous-Batching Serving}
\label{sec:eval-continuous-batching}
We evaluate Qwen3-235B-A22B with 4,096-token prompts, 128 generated tokens per request, \(B_{\max}{=}32\), an 8,192-token iteration budget, and 512-token prefill chunks.
At each iteration boundary, the scheduler reserves one decode token per active request before admitting at most one prefill chunk per waiting request.
Mixed co-schedules decode and prefill in one pass, whereas Serial uses separate decode and prefill passes, reusing each fetched expert across routed rows within a pass.
For each mode, all configurations replay identical arrivals and use the same analytical routing with balanced per-expert rows, scheduling, placement, capacities, and H100 BF16 timings.
DirectOnly carries HBF-resident experts over the two GPU--HBF links, whereas RelayOnly uses the two HBF--HBM--GPU relay routes; \systemname{} uses both route sets concurrently.
At 1.6\,TB/s per link, the aggregate bandwidth for SRAM-ready expert data is up to 3.2\,TB/s in either baseline and 6.4\,TB/s in \systemname{}.
\systemname{} assigns each active expert entirely to either Direct or Relay without splitting its weight tiles across the two.
Because both route sets operate concurrently, expert-delivery time is \(\max(T_{\mathrm{Direct}},T_{\mathrm{Relay}})\).
We replay five Poisson traces at 50\%, 75\%, and 90\% of RelayOnly's backlogged throughput under Mixed, discarding 128 warm-up requests and measuring the next 1,024 requests per trace.
Table~\ref{tab:table_3} summarizes P90 TPOT and P90 E2E latency.

Under Mixed, \systemname{} reduces both P90 metrics in every seed, load, and baseline comparison.
Mixed raises peak throughput over Serial by 10.3\% for \systemname{} and 12.6\% for the baselines, while \systemname{} retains 34.1\%--37.1\% higher peak throughput than the single-route baselines across both modes.

\begin{figure}[!t]
  \centering
  \includegraphics[width=\columnwidth]{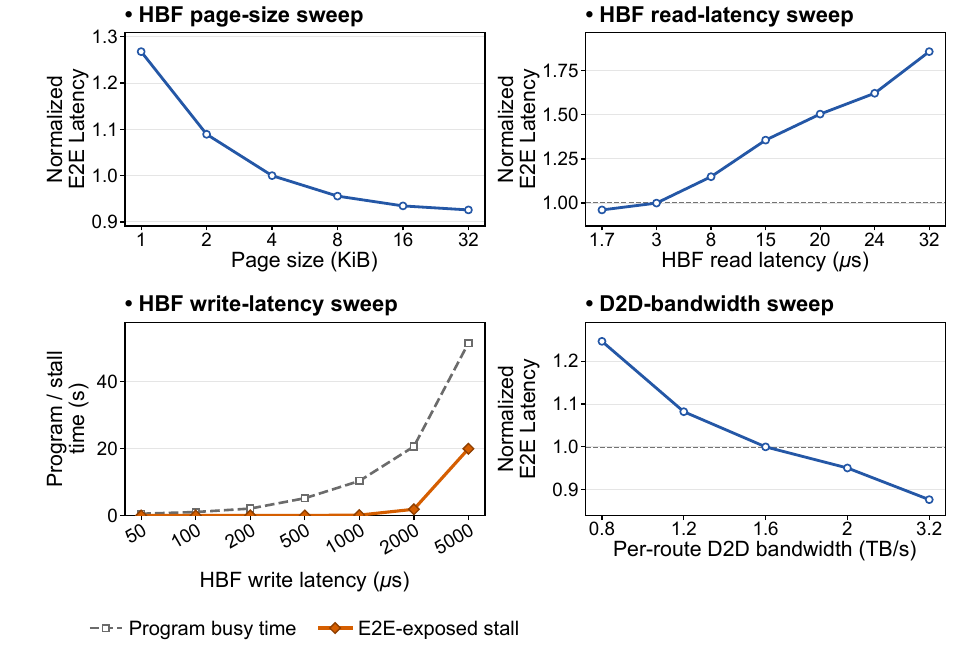}
  \caption{HBF parameter sensitivity. All panels report normalized E2E latency except the program-latency panel, which reports cumulative program-busy and E2E-exposed stall time.}
  \label{fig:fig_12}
\end{figure}
\subsection{Sensitivity Analysis}
\label{sec:sensitivity}

Figure~\ref{fig:fig_12} evaluates the sensitivity of \systemname{} to key HBF parameters using Llama~4 Maverick at $B/L_{\mathrm{in}}/L_{\mathrm{out}}=1/4\mathrm{K}/1\mathrm{M}$.
The sweep uses asymmetric KV placement: the KV state produced by the 4K-token prefill is placed in HBF, whereas newly generated decode KV is appended to HBM.
When the package-wide HBM-resident decode KV reaches the 24.46\,GB decode-KV budget, the oldest page-aligned decode KV begins to spill from HBM to HBF.
At the final context, 24.46\,GB of decode KV is balanced across the two 24-GB HBM stacks (approximately 12.23\,GB per stack), while the remaining 172.96\,GB resides in HBF, for a total KV cache size of 197.41\,GB.
For the page-size sensitivity study only, we scale the transfer-region capacity in proportion to the bytes delivered per page wave.

For HBF page size and read latency, results are normalized to the 4\,KiB and 3\,$\mu$s configurations, respectively.
As shown in Figure~\ref{fig:fig_12}, a larger page size increases the amount of data transferred per HBF access, improving the effective aggregate read bandwidth and reducing E2E latency.
Similarly, lower read latency allows more HBF reads to complete within a given time, further increasing the effective aggregate read bandwidth.
Because \systemname{} concurrently delivers different experts over the Direct and Relay routes, it can better utilize the additional HBF-side data availability enabled by larger transfers or lower read latency, translating device-level improvements into further end-to-end gains.

The HBF write-latency sweep reports total program-busy time, which measures the time spent programming HBF, and E2E-exposed stall, the portion not overlapped with computation or data movement.
\systemname{} hides all program time through 500\,$\mu$s by scheduling writes alongside independent work, so the exposed stall remains zero even as program-busy time increases.
At 5~ms, programming exceeds the available overlap window, adding 19.94\,s of stall over the full execution.
Thus, raw HBF program latency does not directly translate into serving delay; only the unhidden portion affects E2E latency.

For per-route D2D bandwidth, results are normalized to the 1.6\,TB/s configuration, using the published HBF stack-level read-bandwidth target as the D2D-link provisioning point ~\cite{sandisk_hbf}.
NAND-read timing and SRAM readiness remain separately modeled.
E2E latency increases as D2D bandwidth decreases below 1.6\,TB/s, indicating that the link can become a bottleneck after data become ready in SRAM.
Conversely, increasing bandwidth beyond 1.6\,TB/s further reduces E2E latency, demonstrating that \systemname{} can translate additional GPU-facing D2D bandwidth into end-to-end performance gains.

\subsection{Cost Sensitivity and Interconnect Scalability}
\label{sec:cost-interconnect}

Because the public HBF fact sheet reports neither a numerical per-stack price nor package-integration cost~\cite{sandisk_hbf}, we present a parametric nominal-capacity-per-cost analysis of the memory subsystem rather than an absolute cost estimate.
Let $C_{\mathrm{HBM}}$ denote the cost of one 24-GB HBM3E stack, let $r=C_{\mathrm{HBF}}/C_{\mathrm{HBM}}$ denote the HBF-to-HBM per-stack cost ratio, and let $\delta=\Delta C_{\mathrm{int}}/C_{\mathrm{HBM}}\geq 0$ denote \systemname{}-specific incremental integration cost relative to the four-HBM reference, including additional peer PHYs and base-die logic, package integration, and yield loss not already included in the stack costs.
Four 24-GB HBM3E stacks provide 96 GB of nominal capacity, whereas two 24-GB HBM3E stacks and two 512-GB HBF stacks provide 1,072 GB~\cite{micron_hbm3e,sandisk_hbf}.
The resulting reference-normalized nominal-capacity-per-cost gain is $G(r,\delta)=44.67/(2+2r+\delta)$.
We use $r=1$ only as an illustrative anchor for the fact sheet's qualitative similar-cost statement, not as a vendor-quoted price point~\cite{sandisk_hbf}.
The fact sheet reports an HBM4-like footprint and stack height, motivating our equal-stack-count comparison~\cite{sandisk_hbf}.
At the illustrative $(r,\delta)=(1,0)$ anchor, $G=11.17\times$, while $G>1$ whenever $2r+\delta<42.67$, equivalently $r<21.33$ for $\delta=0$.

For $n$ HBM--HBF pairs, \systemname{} uses $2n$ GPU-facing stack connections---the same number as a $2n$-HBM reference---and adds $n$ pair-local HBM--HBF links, yielding $3n$ logical links and $O(n)$ link growth.
Each additional pair adds HBF capacity together with a one-hop direct path and a two-hop pair-local relay path, without centralizing relay traffic or rewiring existing pairs.
In the evaluated four-stack floorplan, facing-edge UCIe-A PHY placement keeps every UCIe mainband channel within the target advanced-package reach limit~\cite{ucie_2mm}, while larger configurations remain subject to GPU-edge PHY beachfront, package area, and interposer routability.

\subsection{HBF Endurance}
\label{sec:hbf-endurance}
We project HBF endurance from event-level logical KV writes generated by the evaluated serving workload, rather than from accelerated-aging measurements.
Following prior work~\cite{hbf_kv_cache}, we assume $E_{\mathrm{SLC}}=100{,}000$ program/erase cycles.
The projected lifetime is
\begin{equation}
  \mathrm{Lifetime} =
  \frac{C_{\mathrm{KV}}E_{\mathrm{SLC}}}
       {A_W R_{\mathrm{HBF}}u},
  \label{eq:hbf-lifetime}
\end{equation}
where $C_{\mathrm{KV}}$ is writable KV capacity, $R_{\mathrm{HBF}}$ is the logical HBF write rate, $A_W$ is the physical-to-logical write-amplification factor (WAF)~\cite{hu2009writeamp}, and $u$ is the duty factor.
Using the request-stateful continuous-batching scheduler described in Section~\ref{sec:eval-continuous-batching}, we evaluate a prefill-heavy, write-stress case with Llama 4 Maverick at $L_{\mathrm{in}}/L_{\mathrm{out}}{=}32\mathrm{K}/128$ and $B_{\max}{=}32$, where long prefills dominate HBF KV writes.
After warm-up, 256 completed requests generate 1,648.34\,GB of logical HBF writes over 1,623.76\,s, yielding $R_{\mathrm{HBF}}{=}1{,}015.13$\,MB/s; every measured iteration contains both prefill and decode work, and the 128- and 256-request estimates differ by only 0.0064\%.
The run remains within the 1,024-GB HBF capacity through capacity-aware admission.
After 801.42\,GB of fixed allocation and a 16-GB reserve for garbage collection, $C_{\mathrm{KV}}{=}206.58$\,GB.
This gives a 0.645-year continuous-activity projection at $A_W{=}1$ and $u{=}1$, assuming uniform wear and no bad-block loss.
All measured HBF writes are page-aligned prefill writes: decode KV remains in HBM, and programmed bytes match logical bytes, giving a modeled page-padding factor of 1.0.
Thus, sustained long-prompt endurance is governed by the rate at which prefill KV cycles through the writable region, rather than nominal HBF capacity; \systemname{} confines this unavoidable traffic to large aligned transfers without adding token-granular decode programs or page-padding overhead.
A deployed-lifetime claim additionally requires target-device measurements of total WAF, per-block erase counts, bad-block growth, and device aging, together with a deployment-specific duty cycle~\cite{meza2015flashfailures,schroeder2016flash}.
\section{Related Work}
\label{sec:related-work}
Prior work has largely treated HBF organization and serving scheduling as separate concerns.
H$^3$ places large read-only data in HBF and mutable state in HBM~\cite{h3}; \systemname{} goes beyond this placement split by serving HBF-resident data over concurrent Direct and Relay paths.
A recent study analyzes the opportunities and challenges of using HBF as GPU memory for LLM inference~\cite{son2026hbf}, while complementary work evaluates HBF-resident KV cache performance and endurance~\cite{hbf_kv_cache}; \systemname{} additionally coordinates latency-critical expert delivery with prefill- and decode-specific KV movement.
An HBM--HBF pooling architecture places HBF behind HBM and uses a custom base die for layer-wise prefetching and buffered writeback~\cite{hbm_hbf_pooling}; \systemname{} instead retains direct GPU--HBF access and uses the HBM relay path concurrently.
Within this topology, \systemname{} distinguishes four execution modes, matching each mode to its bandwidth and access granularity.
Recent expert-prefetching methods predict or speculate future expert selections but do not guarantee the exact top-$k$ decision~\cite{madan2026speculating,zhu2025preattention}; \systemname{} provides exact early selection by reformulation of bias-free router.
Unlike Sarathi-Serve, which coordinates prefill and decode without modeling HBF transfers or program operations~\cite{sarathi_serve}, \systemname{} jointly schedules continuous batching, HBF paths, and writes for dynamic MoE serving.
\section{Conclusion}
\label{sec:conclusion}

This paper presents \systemname{}, a GPU memory architecture that connects HBM and HBF as main memory tiers for LLM inference.
A dedicated HBM--HBF path enables HBF-resident data to reach the GPU either directly or through HBM.
\textsc{\systemname{}} combines this dual-path access with workload-aware data placement, early expert determination, and prefill/decode-aware KV write scheduling to hide HBF access latency while avoiding interference with latency-critical reads.
\systemname{} also addresses the endurance challenges of flash-based memory by managing KV cache writes as part of the serving schedule.
By using HBF as a main memory rather than relying solely on costly HBM capacity, \textsc{\systemname{}} provides a more cost-effective and scalable system to support larger models and longer KV caches.


\bibliographystyle{IEEEtranS}
\bibliography{99_reference}

@article{sharma2022universal,
  title={Universal chiplet interconnect express (UCIe): An open industry standard for innovations with chiplets at package level},
  author={Sharma, Debendra Das and Pasdast, Gerald and Qian, Zhiguo and Aygun, Kemal},
  journal={IEEE Transactions on Components, Packaging and Manufacturing Technology},
  volume={12},
  number={9},
  pages={1423--1431},
  year={2022},
  publisher={IEEE}
}

@inproceedings{he2016deep,
  title={Deep residual learning for image recognition},
  author={He, Kaiming and Zhang, Xiangyu and Ren, Shaoqing and Sun, Jian},
  booktitle={Proceedings of the IEEE conference on computer vision and pattern recognition},
  pages={770--778},
  year={2016}
}

@misc{nvidia_h100_specs,
  author       = {{NVIDIA Corporation}},
  title        = {{NVIDIA H100 Tensor Core GPU Architecture}},
  year         = {2023},
  howpublished = {Architecture whitepaper, version 1.04},
  url          = {https://resources.nvidia.com/en-us-hopper-architecture/nvidia-h100-tensor-c},
  note         = {Tables 1 and 3 report 756 dense and 1,513 sparse {BF16}
                  Tensor {TFLOP/s} with {FP32} accumulation for {H100 PCIe};
                  accessed 2026-07-19}
}

@misc{mixtral8x22b,
  author       = {{Mistral AI}},
  title        = {{Mixtral-8x22B-v0.1} Model Card and Configuration},
  year         = {2024},
  url          = {https://huggingface.co/mistralai/Mixtral-8x22B-v0.1},
  note         = {Accessed: 2026-07-14}
}

@misc{qwen3_235b,
  author       = {{Qwen Team}},
  title        = {{Qwen3-235B-A22B}: Model Announcement and Configuration},
  year         = {2025},
  url          = {https://qwenlm.github.io/blog/qwen3/},
  note         = {Accessed: 2026-07-14}
}

@misc{grok1,
  author       = {{xAI}},
  title        = {{Grok-1} Open-Weight Model Repository},
  year         = {2024},
  url          = {https://github.com/xai-org/grok-1},
  note         = {Accessed: 2026-07-14}
}

@misc{llama4_maverick,
  author       = {{Meta AI}},
  title        = {{Llama 4 Maverick 17B-128E} Model Card and Configuration},
  year         = {2025},
  url          = {https://huggingface.co/meta-llama/Llama-4-Maverick-17B-128E-Instruct},
  note         = {Accessed: 2026-07-14}
}

@article{deepseekv2,
  author  = {{DeepSeek-AI} and Aixin Liu and Bei Feng and Bin Wang and
             Bingxuan Wang and Bo Liu and Chenggang Zhao and Chengqi Dengr and
             Chong Ruan and Damai Dai and Daya Guo and Dejian Yang and
             Deli Chen and Dongjie Ji and Erhang Li and Fangyun Lin and
             Fuli Luo and Guangbo Hao and Guanting Chen and Guowei Li and
             H. Zhang and Hanwei Xu and Hao Yang and Haowei Zhang and
             Honghui Ding and Huajian Xin and Huazuo Gao and Hui Li and Hui Qu and
             J. L. Cai and Jian Liang and Jianzhong Guo and Jiaqi Ni and
             Jiashi Li and Jin Chen and Jingyang Yuan and Junjie Qiu and
             Junxiao Song and Kai Dong and Kaige Gao and Kang Guan and Lean Wang and
             Lecong Zhang and Lei Xu and Leyi Xia and Liang Zhao and Liyue Zhang and
             Meng Li and Miaojun Wang and Mingchuan Zhang and Minghua Zhang and
             Minghui Tang and Mingming Li and Ning Tian and Panpan Huang and
             Peiyi Wang and Peng Zhang and Qihao Zhu and Qinyu Chen and Qiushi Du and
             R. J. Chen and R. L. Jin and Ruiqi Ge and Ruizhe Pan and Runxin Xu and
             Ruyi Chen and S. S. Li and Shanghao Lu and Shangyan Zhou and
             Shanhuang Chen and Shaoqing Wu and Shengfeng Ye and Shirong Ma and
             Shiyu Wang and Shuang Zhou and Shuiping Yu and Shunfeng Zhou and
             Size Zheng and T. Wang and Tian Pei and Tian Yuan and Tianyu Sun and
             W. L. Xiao and Wangding Zeng and Wei An and Wen Liu and Wenfeng Liang and
             Wenjun Gao and Wentao Zhang and X. Q. Li and Xiangyue Jin and
             Xianzu Wang and Xiao Bi and Xiaodong Liu and Xiaohan Wang and
             Xiaojin Shen and Xiaokang Chen and Xiaosha Chen and Xiaotao Nie and
             Xiaowen Sun and Xiaoxiang Wang and Xin Liu and Xin Xie and Xingkai Yu and
             Xinnan Song and Xinyi Zhou and Xinyu Yang and Xuan Lu and Xuecheng Su and
             Y. Wu and Y. K. Li and Y. X. Wei and Y. X. Zhu and Yanhong Xu and
             Yanping Huang and Yao Li and Yao Zhao and Yaofeng Sun and Yaohui Li and
             Yaohui Wang and Yi Zheng and Yichao Zhang and Yiliang Xiong and
             Yilong Zhao and Ying He and Ying Tang and Yishi Piao and Yixin Dong and
             Yixuan Tan and Yiyuan Liu and Yongji Wang and Yongqiang Guo and
             Yuchen Zhu and Yuduan Wang and Yuheng Zou and Yukun Zha and Yunxian Ma and
             Yuting Yan and Yuxiang You and Yuxuan Liu and Z. Z. Ren and Zehui Ren and
             Zhangli Sha and Zhe Fu and Zhen Huang and Zhen Zhang and Zhenda Xie and
             Zhewen Hao and Zhihong Shao and Zhiniu Wen and Zhipeng Xu and
             Zhongyu Zhang and Zhuoshu Li and Zihan Wang and Zihui Gu and Zilin Li and
             Ziwei Xie},
  title   = {{DeepSeek-V2}: A Strong, Economical, and Efficient
             Mixture-of-Experts Language Model},
  journal = {arXiv preprint arXiv:2405.04434},
  year    = {2024},
  url     = {https://arxiv.org/abs/2405.04434}
}

@article{deepseekv3,
  author  = {{DeepSeek-AI} and Aixin Liu and Bei Feng and Bing Xue and
             Bingxuan Wang and Bochao Wu and Chengda Lu and Chenggang Zhao and
             Chengqi Deng and Chenyu Zhang and Chong Ruan and Damai Dai and
             Daya Guo and Dejian Yang and Deli Chen and Dongjie Ji and Erhang Li and
             Fangyun Lin and Fucong Dai and Fuli Luo and Guangbo Hao and
             Guanting Chen and Guowei Li and H. Zhang and Han Bao and Hanwei Xu and
             Haocheng Wang and Haowei Zhang and Honghui Ding and Huajian Xin and
             Huazuo Gao and Hui Li and Hui Qu and J. L. Cai and Jian Liang and
             Jianzhong Guo and Jiaqi Ni and Jiashi Li and Jiawei Wang and Jin Chen and
             Jingchang Chen and Jingyang Yuan and Junjie Qiu and Junlong Li and
             Junxiao Song and Kai Dong and Kai Hu and Kaige Gao and Kang Guan and
             Kexin Huang and Kuai Yu and Lean Wang and Lecong Zhang and Lei Xu and
             Leyi Xia and Liang Zhao and Litong Wang and Liyue Zhang and Meng Li and
             Miaojun Wang and Mingchuan Zhang and Minghua Zhang and Minghui Tang and
             Mingming Li and Ning Tian and Panpan Huang and Peiyi Wang and Peng Zhang and
             Qiancheng Wang and Qihao Zhu and Qinyu Chen and Qiushi Du and R. J. Chen and
             R. L. Jin and Ruiqi Ge and Ruisong Zhang and Ruizhe Pan and Runji Wang and
             Runxin Xu and Ruoyu Zhang and Ruyi Chen and S. S. Li and Shanghao Lu and
             Shangyan Zhou and Shanhuang Chen and Shaoqing Wu and Shengfeng Ye and
             Shirong Ma and Shiyu Wang and Shuang Zhou and Shuiping Yu and
             Shunfeng Zhou and Shuting Pan and T. Wang and Tao Yun and Tian Pei and
             Tianyu Sun and W. L. Xiao and Wangding Zeng and Wanjia Zhao and Wei An and
             Wen Liu and Wenfeng Liang and Wenjun Gao and Wenqin Yu and Wentao Zhang and
             X. Q. Li and Xiangyue Jin and Xianzu Wang and Xiao Bi and Xiaodong Liu and
             Xiaohan Wang and Xiaojin Shen and Xiaokang Chen and Xiaokang Zhang and
             Xiaosha Chen and Xiaotao Nie and Xiaowen Sun and Xiaoxiang Wang and
             Xin Cheng and Xin Liu and Xin Xie and Xingchao Liu and Xingkai Yu and
             Xinnan Song and Xinxia Shan and Xinyi Zhou and Xinyu Yang and Xinyuan Li and
             Xuecheng Su and Xuheng Lin and Y. K. Li and Y. Q. Wang and Y. X. Wei and
             Y. X. Zhu and Yang Zhang and Yanhong Xu and Yanping Huang and
             Yao Li and Yao Zhao and Yaofeng Sun and Yaohui Li and Yaohui Wang and
             Yi Yu and Yi Zheng and Yichao Zhang and Yifan Shi and Yiliang Xiong and
             Ying He and Ying Tang and Yishi Piao and Yisong Wang and Yixuan Tan and
             Yiyang Ma and Yiyuan Liu and Yongqiang Guo and Yu Wu and Yuan Ou and
             Yuchen Zhu and Yuduan Wang and Yue Gong and Yuheng Zou and Yujia He and
             Yukun Zha and Yunfan Xiong and Yunxian Ma and Yuting Yan and Yuxiang Luo and
             Yuxiang You and Yuxuan Liu and Yuyang Zhou and Z. F. Wu and Z. Z. Ren and
             Zehui Ren and Zhangli Sha and Zhe Fu and Zhean Xu and Zhen Huang and
             Zhen Zhang and Zhenda Xie and Zhengyan Zhang and Zhewen Hao and Zhibin Gou and
             Zhicheng Ma and Zhigang Yan and Zhihong Shao and Zhipeng Xu and Zhiyu Wu and
             Zhongyu Zhang and Zhuoshu Li and Zihui Gu and Zijia Zhu and Zijun Liu and
             Zilin Li and Ziwei Xie and Ziyang Song and Ziyi Gao and Zizheng Pan},
  title   = {{DeepSeek-V3} Technical Report},
  journal = {arXiv preprint arXiv:2412.19437},
  year    = {2024},
  url     = {https://arxiv.org/abs/2412.19437}
}

@manual{ucie_spec3,
  author       = {{UCIe Consortium}},
  title        = {Universal Chiplet Interconnect Express
                  ({UCIe}) Specification, Revision 3.0},
  organization = {UCIe Consortium},
  year         = {2025},
  month        = aug,
  note         = {Supports 48- and 64-GT/s data rates;
                  accessed July 29, 2026},
  url          = {https://www.uciexpress.org/specifications}
}

@article{sharma_coughlin2024ucie_memory_storage,
  author  = {Debendra Das Sharma and Thomas M. Coughlin},
  title   = {Universal Chiplet Interconnect Express:
             An Open Industry Standard for Memory and
             Storage Applications},
  journal = {Computer},
  volume  = {57},
  number  = {1},
  pages   = {75--81},
  year    = {2024},
  month   = jan,
  doi     = {10.1109/MC.2023.3318769}
}

@inproceedings{das_sharma2025ucie_memory,
  author       = {Das Sharma, Debendra and Choudhary, Swadesh and
                  Onufryk, Peter and Pelt, Rob},
  title        = {{On-Package Memory with Universal Chiplet Interconnect
                  Express (UCIe): A Low Power, High Bandwidth, Low Latency
                  and Low Cost Approach}},
  booktitle    = {Proceedings of the IEEE Symposium on High-Performance
                  Interconnects (HOTI)},
  year         = {2025},
  pages        = {87--96},
  doi          = {10.1109/HOTI66940.2025.00027},
  url          = {https://arxiv.org/abs/2510.06513}
}

@misc{sandisk_hbf_patent_2025,
  author       = {Xiang Yang and Deepanshu Dutta and Yan Li and
                  Masaaki Higashitani},
  title        = {High Bandwidth Nonvolatile Memory Devices},
  year         = {2025},
  howpublished = {U.S. Patent Application Publication US 2025/0259685 A1,
                  assigned to SanDisk Technologies LLC},
  url          = {https://patents.google.com/patent/US20250259685A1/en},
  note         = {Accessed: 2026-07-18}
}

@inproceedings{yu2022orca,
  author    = {Gyeong-In Yu and Joo Seong Jeong and Geon-Woo Kim and
               Soojeong Kim and Byung-Gon Chun},
  title     = {{Orca}: A Distributed Serving System for
               Transformer-Based Generative Models},
  booktitle = {16th USENIX Symposium on Operating Systems Design and
               Implementation (OSDI)},
  pages     = {521--538},
  year      = {2022},
  url       = {https://www.usenix.org/conference/osdi22/presentation/yu}
}

@inproceedings{narayanan2021megatron,
  author    = {Deepak Narayanan and Mohammad Shoeybi and Jared Casper and
               Patrick LeGresley and Mostofa Patwary and Vijay Anand Korthikanti
               and Dmitri Vainbrand and Prethvi Kashinkunti and Julie Bernauer
               and Bryan Catanzaro and Amar Phanishayee and Matei Zaharia},
  title     = {Efficient Large-Scale Language Model Training on {GPU} Clusters
               Using {Megatron-LM}},
  booktitle = {Proceedings of the International Conference for High Performance
               Computing, Networking, Storage and Analysis (SC)},
  year      = {2021},
  doi       = {10.1145/3458817.3476209},
  url       = {https://cs.stanford.edu/people/matei/papers/2021/sc_megatron_lm.pdf}
}

@inproceedings{sarathi_serve,
  author    = {Amey Agrawal and Nitin Kedia and Ashish Panwar and Jayashree
               Mohan and Nipun Kwatra and Bhargav S. Gulavani and
               Alexey Tumanov and Ramachandran Ramjee},
  title     = {Taming Throughput-Latency Tradeoff in {LLM} Inference with
               {Sarathi-Serve}},
  booktitle = {18th USENIX Symposium on Operating Systems Design and
               Implementation (OSDI)},
  pages     = {117--134},
  year      = {2024}
}

@article{h3,
  title={H 3: Hybrid architecture using high bandwidth memory and high bandwidth flash for cost-efficient LLM inference},
  author={Ha, Minho and Kim, Euiseok and Kim, Hoshik},
  journal={IEEE Computer Architecture Letters},
  year={2026},
  publisher={IEEE}
}

@article{hbm_hbf_pooling,
  title={HBM-HBF-Centric Memory Pooling Architecture With Custom Base Die for Terabyte-Scale LLM Inference},
  author={Park, Junho and An, Hyowon and Suh, Haeseok and Yoon, Youngsu and Lee, Hyuni and Kim, Joungho},
  journal={IEEE Computer Architecture Letters},
  year={2026},
  publisher={IEEE}
}

@article{MoE,
  title={Adaptive mixtures of local experts},
  author={Jacobs, Robert A and Jordan, Michael I and Nowlan, Steven J and Hinton, Geoffrey E},
  journal={Neural computation},
  volume={3},
  number={1},
  pages={79--87},
  year={1991},
  publisher={MIT Press}
}

@misc{nvidia_h100,
  title={NVIDIA H100 Tensor Core GPU},
  author={{NVIDIA}},
  year={2024},
  howpublished={\url{https://www.nvidia.com/en-us/data-center/h100/}},
  note={Accessed: 2026-07-21}
}

@article{hbf_kv_cache,
  author  = {Kyung, Kwanhee and Moon, Yeon Ji and Cho, Juhwan and Ahn, Jung Ho},
  title   = {High-Bandwidth Flash for {KV} Caches: Endurance and Performance Implications},
  journal = {IEEE Computer Architecture Letters},
  volume  = {25},
  number  = {1},
  pages   = {210--213},
  year    = {2026},
  month   = jan,
  doi     = {10.1109/LCA.2026.3695938}
}

@manual{kioxia_tc58_2019,
  author       = {{KIOXIA Corporation}},
  title        = {{TC58CYG2S0HRAIJ: 4Gb 1.8V Serial Interface NAND
                   Technical Data Sheet}},
  organization = {KIOXIA Corporation},
  year         = {2019},
  month        = oct,
  url          = {https://europe.kioxia.com/content/dam/kioxia/newidr/productinfo/datasheet/201910/DST_TC58CYG2S0HRAIJ-TDE_EN_36007.pdf},
  note         = {Revision 2.00; programming, reading, and erasing
                  characteristics in Section~3.7; accessed 2026-07-24}
}

@inproceedings{gqa,
  author    = {Joshua Ainslie and James Lee-Thorp and Michiel de Jong and Yury Zemlyanskiy and Federico Lebr{\'o}n and Sumit Sanghai},
  title     = {{GQA}: Training Generalized Multi-Query Transformer Models from Multi-Head Checkpoints},
  booktitle = {Proceedings of the 2023 Conference on Empirical Methods in Natural Language Processing},
  pages     = {4895--4901},
  address   = {Singapore},
  publisher = {Association for Computational Linguistics},
  month     = dec,
  year      = {2023},
  doi       = {10.18653/v1/2023.emnlp-main.298},
  url       = {https://aclanthology.org/2023.emnlp-main.298/}
}

@article{mqa,
  author        = {Noam Shazeer},
  title         = {Fast Transformer Decoding: One Write-Head is All You Need},
  journal       = {arXiv preprint arXiv:1911.02150},
  year          = {2019},
  eprint        = {1911.02150},
  archivePrefix = {arXiv},
  primaryClass  = {cs.NE},
  url           = {https://arxiv.org/abs/1911.02150}
}

@misc{nvlink_c2c,
  author       = {{NVIDIA Corporation}},
  title        = {{NVIDIA Grace Hopper Superchip Architecture In-Depth}},
  year         = {2022},
  month        = nov,
  url          = {https://developer.nvidia.com/blog/nvidia-grace-hopper-superchip-architecture-in-depth/},
  note         = {Accessed: 2026-07-24}
}

@misc{b200,
  author       = {{NVIDIA Corporation}},
  title        = {{NVIDIA HGX B200} Reduces Embodied Carbon Emissions Intensity},
  year         = {2025},
  month        = sep,
  howpublished = {NVIDIA Technical Blog},
  url          = {https://developer.nvidia.com/blog/nvidia-hgx-b200-reduces-embodied-carbon-emissions-intensity/},
  note         = {Reports 180 GB of HBM3e memory per NVIDIA B200 GPU;
                  accessed 2026-07-24}
}

@misc{deepseekv4,
  author       = {{DeepSeek-AI}},
  title        = {{DeepSeek-V4-Pro} Model Card},
  year         = {2026},
  howpublished = {Hugging Face},
  url          = {https://huggingface.co/deepseek-ai/DeepSeek-V4-Pro},
  note         = {Accessed: 2026-07-28}
}

@misc{glm5,
  author       = {{Z.ai}},
  title        = {{GLM-5} Model Card},
  year         = {2026},
  howpublished = {Hugging Face},
  url          = {https://huggingface.co/zai-org/GLM-5},
  note         = {Accessed: 2026-07-28}
}

@article{kimi_k2,
  author  = {{Kimi Team}},
  title   = {{Kimi K2}: Open Agentic Intelligence},
  journal = {arXiv preprint arXiv:2507.20534},
  year    = {2025},
  url     = {https://arxiv.org/abs/2507.20534}
}

@inproceedings{shazeer2017moe,
  author    = {Noam Shazeer and Azalia Mirhoseini and Krzysztof Maziarz and
               Andy Davis and Quoc V. Le and Geoffrey E. Hinton and Jeff Dean},
  title     = {Outrageously Large Neural Networks:
               The Sparsely-Gated Mixture-of-Experts Layer},
  booktitle = {International Conference on Learning Representations},
  year      = {2017},
  url       = {https://openreview.net/forum?id=B1ckMDqlg}
}

@article{fedus2022switch,
  author  = {William Fedus and Barret Zoph and Noam Shazeer},
  title   = {Switch Transformers: Scaling to Trillion Parameter Models
             with Simple and Efficient Sparsity},
  journal = {Journal of Machine Learning Research},
  volume  = {23},
  number  = {120},
  pages   = {1--39},
  year    = {2022},
  url     = {https://jmlr.org/papers/v23/21-0998.html}
}

@inproceedings{agrawal2008ssd,
  author    = {Nitin Agrawal and Vijayan Prabhakaran and Ted Wobber and
               John D. Davis and Mark Manasse and Rina Panigrahy},
  title     = {Design Tradeoffs for {SSD} Performance},
  booktitle = {2008 USENIX Annual Technical Conference (USENIX ATC 08)},
  pages     = {57--70},
  publisher = {USENIX Association},
  year      = {2008},
  url       = {https://www.usenix.org/conference/2008-usenix-annual-technical-conference/design-tradeoffs-ssd-performance}
}

@inproceedings{vaswani2017attention,
  author    = {Ashish Vaswani and Noam Shazeer and Niki Parmar and
               Jakob Uszkoreit and Llion Jones and Aidan N. Gomez and
               Lukasz Kaiser and Illia Polosukhin},
  title     = {Attention Is All You Need},
  booktitle = {Advances in Neural Information Processing Systems},
  volume    = {30},
  year      = {2017},
  url       = {https://proceedings.neurips.cc/paper/7181-attention-is-all-you-need}
}

@inproceedings{pope2023scaling,
  author    = {Reiner Pope and Sholto Douglas and Aakanksha Chowdhery and
               Jacob Devlin and James Bradbury and Anselm Levskaya and
               Jonathan Heek and Kefan Xiao and Shivani Agrawal and Jeff Dean},
  title     = {Efficiently Scaling Transformer Inference},
  booktitle = {Proceedings of Machine Learning and Systems},
  volume    = {5},
  year      = {2023},
  url       = {https://proceedings.mlsys.org/paper_files/paper/2023/hash/c4be71ab8d24cdfb45e3d06dbfca2780-Abstract-mlsys2023.html}
}

@inproceedings{hong2022guardederase,
  author    = {Duwon Hong and Myungsuk Kim and Geonhee Cho and
               Dusol Lee and Jihong Kim},
  title     = {{GuardedErase}: Extending {SSD} Lifetimes by Protecting
               Weak Wordlines},
  booktitle = {20th USENIX Conference on File and Storage Technologies
               (FAST 22)},
  pages     = {133--146},
  publisher = {USENIX Association},
  year      = {2022},
  url       = {https://www.usenix.org/conference/fast22/presentation/hong}
}

@inproceedings{rmsnorm2019,
  author    = {Biao Zhang and Rico Sennrich},
  title     = {Root Mean Square Layer Normalization},
  booktitle = {Advances in Neural Information Processing Systems},
  volume    = {32},
  pages     = {12360--12371},
  year      = {2019},
  url       = {https://proceedings.neurips.cc/paper/2019/hash/1e8a19426224ca89e83cef47f1e7f53b-Abstract.html}
}

@inproceedings{hwang2023tutel,
  author    = {Changho Hwang and Wei Cui and Yifan Xiong and Ziyue Yang
               and Ze Liu and Han Hu and Zilong Wang and Rafael Salas and
               Jithin Jose and Prabhat Ram and Joe Chau and Peng Cheng and
               Fan Yang and Mao Yang and Yongqiang Xiong},
  title     = {Tutel: Adaptive Mixture-of-Experts at Scale},
  booktitle = {Proceedings of Machine Learning and Systems (MLSys)},
  volume    = {5},
  year      = {2023},
  url       = {https://proceedings.mlsys.org/paper_files/paper/2023/hash/5616d34cf8ff73942cfd5aa922842556-Abstract-mlsys2023.html}
}

@article{xue2024moeinfinity,
  author  = {Leyang Xue and Yao Fu and Zhan Lu and Luo Mai and Mahesh Marina},
  title   = {{MoE-Infinity}: Offloading-Efficient {MoE} Model Serving},
  journal = {arXiv preprint arXiv:2401.14361},
  year    = {2024},
  url     = {https://arxiv.org/abs/2401.14361}
}

@inproceedings{kamahori2025fiddler,
  author    = {Keisuke Kamahori and Tian Tang and Yile Gu and Kan Zhu and
               Baris Kasikci},
  title     = {Fiddler: CPU--GPU Orchestration for Fast Inference of
               Mixture-of-Experts Models},
  booktitle = {International Conference on Learning Representations},
  year      = {2025},
  url       = {https://openreview.net/forum?id=N5fVv6PZGz}
}

@misc{intel8452y,
  author       = {{Intel Corporation}},
  title        = {Intel Xeon Platinum 8452Y Processor Specifications},
  year         = {2023},
  howpublished = {\url{https://www.intel.com/content/www/us/en/products/sku/231761/intel-xeon-platinum-8452y-processor-67-5m-cache-2-00-ghz/specifications.html}},
  note         = {Accessed July 28, 2026}
}

@article{ucie_sip_2024,
  author  = {Debendra Das Sharma and Gerald Pasdast and Sathya Tiagaraj
             and Kemal Ayg{\"u}n},
  title   = {High-Performance, Power-Efficient Three-Dimensional
             System-in-Package Designs with Universal Chiplet
             Interconnect Express},
  journal = {Nature Electronics},
  volume  = {7},
  pages   = {244--254},
  year    = {2024},
  doi     = {10.1038/s41928-024-01126-y}
}

@article{wu_guo2025ucie64g,
  author  = {Zuoguo Wu and Jong-Ru Guo},
  title   = {Analysis of {UCIe} 48/64-{GT/s} Electrical Links},
  journal = {IEEE Open Journal of the Solid-State Circuits Society},
  volume  = {5},
  pages   = {401--409},
  year    = {2025},
  doi     = {10.1109/OJSSCS.2025.3605558}
}

@article{zhu2025preattention,
  author  = {Shien Zhu and Samuel Bohl and Robin Oester and Gustavo Alonso},
  title   = {Pre-Attention Expert Prediction and Prefetching for
             Mixture-of-Experts Large Language Models},
  journal = {arXiv preprint arXiv:2511.10676},
  year    = {2025},
  doi     = {10.48550/arXiv.2511.10676},
  url     = {https://arxiv.org/abs/2511.10676}
}

@article{madan2026speculating,
  author  = {Vivan Madan and Prajwal Singhania and Abhinav Bhatele and
             Tom Goldstein and Ashwinee Panda},
  title   = {Speculating Experts Accelerates Inference for
             Mixture-of-Experts},
  journal = {arXiv preprint arXiv:2603.19289},
  year    = {2026},
  doi     = {10.48550/arXiv.2603.19289},
  url     = {https://arxiv.org/abs/2603.19289}
}

@article{jiang2024mixtral,
  author  = {Albert Q. Jiang and
             Alexandre Sablayrolles and
             Antoine Roux and
             Arthur Mensch and
             Blanche Savary and
             Chris Bamford and
             Devendra Singh Chaplot and
             Diego de las Casas and
             Emma {Bou Hanna} and
             Florian Bressand and
             Gianna Lengyel and
             Guillaume Bour and
             Guillaume Lample and
             L{\'e}lio {Renard Lavaud} and
             Lucile Saulnier and
             Marie-Anne Lachaux and
             Pierre Stock and
             Sandeep Subramanian and
             Sophia Yang and
             Szymon Antoniak and
             Teven {Le Scao} and
             Th{\'e}ophile Gervet and
             Thibaut Lavril and
             Thomas Wang and
             Timoth{\'e}e Lacroix and
             William {El Sayed}},
  title   = {Mixtral of Experts},
  journal = {arXiv preprint arXiv:2401.04088},
  year    = {2024},
  doi     = {10.48550/arXiv.2401.04088},
  url     = {https://arxiv.org/abs/2401.04088}
}

@inproceedings{huang2024moeinference,
  author    = {Haiyang Huang and Newsha Ardalani and Anna Sun and Liu Ke and
               Hsien-Hsin S. Lee and Shruti Bhosale and
               Carole-Jean Wu and Benjamin Lee},
  title     = {Toward Efficient Inference for Mixture of Experts},
  booktitle = {Advances in Neural Information Processing Systems},
  volume    = {37},
  pages     = {84033--84059},
  year      = {2024},
  doi       = {10.52202/079017-2670},
  url       = {https://proceedings.neurips.cc/paper_files/paper/2024/hash/98bf3b8505c611ac21055dd9d355c66e-Abstract-Conference.html}
}

@techreport{micron_hbm3e,
  author      = {{Micron Technology, Inc.}},
  title       = {{Micron HBM3E Product Brief}},
  institution = {Micron Technology, Inc.},
  type        = {Product Brief},
  number      = {Rev. C},
  month       = oct,
  year        = {2023},
  url         = {https://assets.micron.com/adobe/assets/urn\%3Aaaid\%3Aaem\%3Ab710d8f2-7f66-44c1-a234-456e2b986347/renditions/original/as/hbm3e-product-brief.pdf},
  note        = {Accessed: Jul. 31, 2026}
}

@techreport{sandisk_hbf,
  author      = {{SanDisk Corporation}},
  title       = {{High Bandwidth Flash (HBF) Fact Sheet}},
  institution = {SanDisk Corporation},
  type        = {Tech Brief},
  month       = jul,
  year        = {2025},
  url         = {https://documents.sandisk.com/content/dam/asset-library/en_us/assets/public/sandisk/collateral/company/Sandisk-HBF-Fact-Sheet.pdf},
  note        = {Tech brief; first-generation target of 512~GB capacity and
                 1.6~TB/s read bandwidth per 16-die stack; accessed
                 Jul. 31, 2026}
}

@misc{ucie_2mm,
  author       = {Stefan Rusu},
  title        = {How {UCIe} Will Enable Broad Chiplet Adoption},
  howpublished = {{UCIe Consortium}},
  month        = mar,
  year         = {2023},
  url          = {https://www.uciexpress.org/post/how-ucie-will-enable-broad-chiplet-adoption},
  note         = {Describes silicon-interposer and bridge implementations
                  of UCIe advanced packaging with channels up to 2 mm;
                  accessed Jul. 31, 2026}
}

@article{son2026hbf,
  author  = {Dowon Son and Yonggon Park and Hyunuk Cho and Hyungkyu Ham and Onur Mutlu and Sungjin Lee and Gwangsun Kim and Jisung Park},
  title   = {{Exploring High-Bandwidth Flash for Modern LLM Inference: Opportunities and Challenges}},
  journal = {IEEE Computer Architecture Letters},
  volume  = {25},
  number  = {2},
  pages   = {251--254},
  year    = {2026},
  doi     = {10.1109/LCA.2026.3705817}
}

@misc{llmsim,
  author       = {{SCALE Lab, Seoul National University}},
  title        = {{LLMSimulator}},
  year         = {2025},
  month        = oct,
  howpublished = {GitHub repository},
  url          = {https://github.com/scale-snu/LLMSimulator},
  note         = {Git commit 419252761fbdb95b789778a02256d458a5537ec7;
                  accessed July 31, 2026}
}

@inproceedings{hu2009writeamp,
  author    = {Xiao-Yu Hu and Evangelos Eleftheriou and Robert Haas
               and Ilias Iliadis and Roman Pletka},
  title     = {Write Amplification Analysis in Flash-Based Solid State Drives},
  booktitle = {Proceedings of SYSTOR 2009: The Israeli Experimental
               Systems Conference},
  pages     = {10:1--10:9},
  year      = {2009},
  publisher = {ACM},
  doi       = {10.1145/1534530.1534544}
}

@inproceedings{schroeder2016flash,
  author    = {Bianca Schroeder and Raghav Lagisetty and Arif Merchant},
  title     = {Flash Reliability in Production:
               The Expected and the Unexpected},
  booktitle = {14th USENIX Conference on File and Storage Technologies
               ({FAST} 16)},
  pages     = {67--80},
  year      = {2016},
  publisher = {USENIX Association}
}

@manual{h100_pcie,
  author       = {{NVIDIA}},
  title        = {{NVIDIA H100 PCIe GPU Product Brief}},
  organization = {{NVIDIA Corporation}},
  number       = {PB-11133-001_v02},
  year         = {2022},
  url          = {https://www.nvidia.com/content/dam/en-zz/Solutions/gtcs22/data-center/h100/PB-11133-001_v01.pdf}
}

@inproceedings{meza2015flashfailures,
  author    = {Justin Meza and Qiang Wu and Sanjeev Kumar and Onur Mutlu},
  title     = {A Large-Scale Study of Flash Memory Failures in the Field},
  booktitle = {Proceedings of the 2015 ACM SIGMETRICS International
               Conference on Measurement and Modeling of Computer Systems},
  pages     = {177--190},
  publisher = {ACM},
  year      = {2015},
  doi       = {10.1145/2745844.2745848},
  url       = {https://doi.org/10.1145/2745844.2745848}
}

\end{document}